\documentclass[11pt,a4paper]{article} 
\pdfoutput=1
\usepackage{epsfig,amssymb,amsmath,latexsym} 
\usepackage{jcappubmy} 
\usepackage{hyperref}  
\usepackage{pifont} 
 
\usepackage{color} 
\definecolor{nicered}{rgb}{0.7,0.1,0.1} 
\definecolor{nicegreen}{rgb}{0.1,0.5,0.1}

\newcommand{\be}{\begin{equation}} 
\newcommand{\ee}{\end{equation}} 
\newcommand{\bea}{\begin{eqnarray}} 
\newcommand{\eea}{\end{eqnarray}}

\newcommand{\de}{\partial}

\renewcommand\o{\omega} 
\renewcommand\a{\alpha} 
 
\renewcommand\b{\beta} 
\renewcommand\k{\kappa} 
\newcommand\e{\epsilon} 
\newcommand\m{\mu} 
\newcommand\n{\nu} 
\newcommand\g{\gamma}

\newcommand\ba{\begin{array}} 
\newcommand\ea{\end{array}}

 \title{ A space dependent Cosmological Constant% for an adiabatic Perfect Fluid 
 } 
 
\author[]{ {\Huge } D. Comelli }
\affiliation[]{INFN - Sezione di Ferrara, Via Saragat 1 I-44122 Ferrara, Italy}
 
\emailAdd{comelli@fe.infn.it}
\date{\small \today} 
 
\abstract{  
In a specific adiabatic perfect fluid, 
 intrinsic entropy density perturbations are 
  the source of   a {\it space-dependent} cosmological constant responsible for  local void inhomogeneity.
    Assuming an anisotropic Locally Rotationally Symmetric space time,
using the 1+1+2 covariant approach and   a Lema$\hat{\rm i}$tre space time metric, we study the cosmological implication of such a scenario giving a proper solution to the Hubble constant tension and providing, locally, also an effective equation of state with $w\leq-1$.
  }

\begin{document}

\maketitle 
 
%\section{Introduction}   
%%%%%%%%%%%%%%%%%%%%%
% \newpage
 %%%%%%%%%%%%%%%%%%%%%
 \section{Introduction}
 
The presence of a Dark Fluid (DF) component in the composition of the Universe is posing challenging problems for his interpretation both to particle physicists than to cosmologists.
The  $\Lambda$CDM model is the simplest model describing the main feature of the present observations \cite{Aghanim:2018eyx} 
 where   
the Dark side of the Universe is given  by a Dark Matter (DM) component (with  zero equation of state $w$ and sound velocity $ c_s^2$) and a Dark Energy (DE) component (with equation of state $w=-1$) consisting of a Cosmological Constant (CC). 
\\
Here we study a   minimal DF model where the late period of the universe, dominated by  DM and DE, is described as a unique adiabatic perfect fluid.
For convenience we name  such a  fluid  Next to Minimal $\Lambda$CDM: N$\Lambda$CDM,
due to the fact that the background dynamics on a Friedman-Robertson-Walker (FRW) space time (without perturbations)   is identical to   the $\Lambda$CDM model.
{ The  N$\Lambda$CDM is described both with a
 {\it field theory} approach, using an EFT  where a single Lagrangian is describing the system and the dof are the goldstone modes of the spontaneously broken symmetries,
  that a {\it fluid description} where only the hydrodynamical eqs take places (conservation eqs of the energy momentum tensor and of some currents).
The first approach is very powerful  to build  specific  Lagrangians  and to check if any symmetry principle is involved into the system (chapter \ref{sect:action}). 
The eqs of motion of the goldstone mode, can be translated in the fluid description that results much easier to be solved and to be compared with physical predictions  (chapters (\ref{ch:NLCDM},\ref{ch:112})).  }
%The  N$\Lambda$CDM is   an effective field theory (chapter (\ref{sect:action}))   describing the dynamics of the goldstone mode
%of the fluid once the equation of state of the system is fixed (given by  the Lagrangian structure).
In particular the N$\Lambda$CDM model results %a  Lagrangian model that corresponds to 
an  adiabatic perfect fluid where
%It represents As well known, entropic perfect fluids conserve
 the ratio of entropy density $s(t,\vec x)$ over number density  $n(t,\vec x)$ (the   entropy per particle $\sigma=s(t,\vec x)/n(t,\vec x)$)   is conserved in time but it results 3-$d$ space dependent 
 $\sigma=\sigma(\vec x)$ (see chapter (\ref{ch:NLCDM})).
 For a particular class of Lagrangians, the entropy per particle is dynamically connected to the CC so that,   the main dynamical implication is the presence of a {\it space dependent CC}. Contrary to the  large amount of literature dedicate to a {\it time dependent CC} (see \cite{Peebles} for a seminal paper and \cite{Macedo} for a recent paper), 
 a space dependent CC   results not to have been explored in depth in  the literature (see for example \cite{Narlikar} for spherically symmetric static solutions).
Analysing  the cosmological perturbations of the N$\Lambda$CDM  model around FRW   it was shown the presence of peculiar growing modes for the comoving curvature perturbations at late time  \cite{Celoria:2017idi}, \cite{Celoria:2017bbh}.\footnote{The model was already partially analysed  in \cite{Celoria:2017xos} where  intrinsic entropy perturbations where   studied.}
%A deeper inspection shows that  the late time instability of the FRW background has 
%his origin  in the  intrinsic  entropy background field   that  sources growing modes for the acceleration field.
We show as   the acceleration (a vector field) of the fundamental observer  is the   source of  the intrinsic instability of the  FRW background (we used the 1+3 covariant approach \cite{Ellis:1998ct}).
      To evade such a behaviour we included the acceleration field into a new background utilising a Locally Rotationally Invariant (LRS) formalism (we used the 1+1+2 covariant approach \cite{vanElst:1995eg}).
      % to   provide the background equations of motion for a system with a fundamental background space-like vector field (the acceleration) (chapter (\ref{ch:112})).
We show as such a DF, also once is dominated by his CC component, doesn't  evolve to a de Sitter phase due to the presence of pressure gradients. % in between the DE and DM components.
The adiabatic fluid contains frozen initial entropic fluctuations that behave effectively as  a space-dependent CC. 
%We studied such a system only during the DM-DE domination period.
In chapter (\ref{ng}) we give the eqs needed to study the propagation of photons in such a space time.
In particular we stress the presence of two Hubble rates: a radial one (along the line of sight) and the orthogonal one (transverse).
In chapter (\ref{ca}) we formulate the evolution equations of the system using a 
Lema$\hat{i}$tre metric.
We perturbatively solve the  eqs of motion in two different setting:
 in chapter (\ref{chFRW}), we start from a FRW background and we add a small $r$-dependent CC as a perturbation while in chapter (\ref{chY}) we obtain an exact small $r$ expansion solution.
In chapter (\ref{geo}) we get the null geodesic solutions   and then
we perform some comparison with the existing nearby cosmological data.
The price to pay for such an unusual spherically symmetric background come from
the constraints   from the observed  homogeneity of the space.
The observer must be positioned close to the centre  of a spherical structure.
 In fact,  from CMB constraint, the observer can be at most few tens of $Mpc$ from the origin in a  Lema$\hat{i}$tre Tolman Bond (LTB void)  \cite{Alnes:2006pf}. 
\footnote{ Strictly speaking, this corresponds to an explicit
violation of the Copernican principle generating  a new fine tuning problem of the order 
$(40\;Mpc/15\;Gpc)^2\sim 10^{-8}$   \cite{Clarkson:2012bg}.
For a recent review about the Copernican principle and the tensions within  the $\Lambda$CDM model see \cite{cea}
%Note that recent observations indicate a 4.9 $\sigma$ tension between the CMB and quasar dipoles
%\cite{Secrest:2020has}.
}

 \section{Effective field theory for Perfect Fluids  and Thermodynamics}
\label{sect:action}
Perfect fluids, and media  in general, can be described, by using an  effective
field theory formalism, in terms of four scalar fields $\Phi^A$  ($A=0,1,2,3$) \cite{carter}. 
\footnote{ Note that a field formulation (with the definition of a Lagrangian) of the dynamics of a media is equivalent to his fluid formulation (equation of motion of  physical quantities as density, pressure, temperature, etc.). The two approaches, in any case, result complementary. The first one is much more powerful once we want stress some symmetry principle, couple exactly the media with gravity or electromagnetism or try some quantum treatment. 
The second one is more physical and intuitive having physical quantities as dynamical variables. For example, the relativistic Kelvin Helmholtz theorem (obtained from the eqs of motion  for perfect fluid) related to 
the vorticity conservation,   in the field formulation  is generated by the Noether’s theorem for the local currents of the volume preserving symmetry. }
The mechanical and thermodynamical properties of the medium
 can be encoded  in a set of  symmetries of the scalar field
action  selecting order by order in a derivative expansion a finite
number of operators.
Following~\cite{Ballesteros:2016gwc} we require the Lagrangian to be invariant under:
\be\label{fields}
\begin{split}
\text{Global shift symmetries:}&  \quad \Phi^A\to \Phi^A+f^A \, ; \\
\text{Field dependent symmetries:} &\quad \Phi^0\to \Phi^0+f(\Phi^a);\quad
\Phi^a\to f^a(\Phi^b),\quad \det\left|\frac{\partial
    f^a}{\partial\Phi^b}\right|=1 \, ;\\
\text{Internal rotational invariance:}& \quad \Phi^a \to {\cal M}^a_b\;\Phi^b \, , \quad {\cal M} \in \text{SO}(3), \;\;a,b=1,2,3 .  \end{split} 
\ee
The global shift symmetry requires the scalars to enter the action
only through their derivatives,  while field dependent symmetries plus
internal rotational invariance select, at the  leading order derivative,  the following operators
\bea
b=\left[\det \left(g^{\mu \nu} \partial_\mu  \Phi^a\,\partial_\nu
    \Phi^b \right) \right]^{1/2},\quad
u^\mu=\frac{1}{b\;\sqrt{g} }\epsilon^{\mu\alpha \beta \gamma}\;\partial_\alpha\Phi^1\;\partial_\beta\Phi^2\;\partial_\gamma\Phi^3,\quad
Y=u^\mu\;\partial_\mu\Phi^0 \,   
\label{shift}
\eea
with $u^2\equiv u^\mu u_\mu = -1$.
As a result, our starting point is the action 
\be\label{actS}
S=\int d^4x\;\sqrt{g}\;{\bf U}(b,\,Y) \, ;
\ee
The corresponding EMT   can be easily obtained using the   formulas
\be
\frac{\delta\, Y}{\delta g^{\m\n}}=-\frac{Y}{2}\;u_\m\;u_\n,\quad 
\frac{\delta\, b}{\delta g^{\m\n}}=\frac{b}{2}\;\left(g_{\mu\nu}+u_\m\;u_\n\right)
%\;\;\to\;\;\frac{\delta Z}{\delta g^{\m\n}}=\frac{Z}{2}\;g_{\m\n}
\ee
that gives the    conserved energy momentum tensor (EMT) 
\be
T_{\mu\nu}=\rho \; u_\mu\;u_\nu+(\rho + p) \; g_{\mu\nu} \, ,
\label{emt}
\ee
that has   a perfect fluid form~\footnote{We have introduced the notation $\frac{\de {\bf U}}{\de Y}
  ={\bf U}_Y$, $\frac{\de {\bf U}}{\de b}
  ={\bf U}_b$ and $\frac{\de^2 {\bf U}}{\de Y\de Y}
  ={\bf U}_{Y^2}$, etc.}  with energy density $\rho$ and pressure $p$ given by
\bea\label{rp}
\rho=-{\bf U}+Y\;{\bf U}_Y,\qquad
p={\bf U}-b\;{\bf U}_b \, .
\eea
The gradient of the velocity field $u_\m$ can be decomposed as (see { appendix A} for   details)
\be\label{gradu}
\nabla_\a\;u_\b=\sigma_{\a\b}+\omega_{\b\a}+\frac{\theta}{3}\;h_{\a\b}-u_\a\;{\cal A}_\b
\ee
where  the tensor   $h_\mu^\nu=\delta_\mu^\nu+u_\mu\,u^\nu$ is a projector on the orthogonal surface to $u^\a$,
$\theta $ is the expansion rate, $\sigma_{\a\b}$ the shear tensor, $\omega_{\a\b}$ the rotation tensor and
${\cal A}_\a=u^\b\nabla_\b u_\a$ the acceleration vector.
\\
We introduce also the   covariant  derivative of a scalar function $f$ along and orthogonal to $u^\m$:
\be
\dot f \equiv u^\b\nabla_\b f,\qquad D_\a f \equiv h_\a^\b\;\nabla_\b f\;\;\;\to\;\;\; \nabla_\a f=-\dot f \;u_\a+D_\a f
\ee
The model features the presence of two
conserved currents
%~\footnote{Actually there more conserved currents
  %that will not be needed here; see~\cite{Ballesteros:2016gwc} for a complete analysis.} 
 (the last one related to the shift symmetry (\ref{shift}) of the $\Phi^0$ field) 
\bea\label{trt}
&& %\nabla^\mu J_\mu = 
\nabla^\mu  (b\,u_\mu)=\dot b+\theta\;b=0 \, ;\\ \label{trt1}&&
% \nabla^\mu {\cal J}_\mu = 
 \nabla^\mu  ({\bf U}_Y\,u_\mu)=\dot {\bf U}_Y+\theta\;{\bf U}_Y=0
 \, ;
\eea
It also follows from (\ref{trt} - \ref{trt1}) that  the ratio
$\sigma\equiv \frac{{\bf U}_Y}{b}$ is conserved, indeed
\be\label{sc}
u^\mu\nabla_\mu\,\sigma\equiv\dot \sigma=0 \, .
\ee
%
%The equations (\ref{trt} - \ref{trt1}) are equivalent to the projection of
%EMT conservation \ref{emt} along $u^\mu$. 
  EMT conservation %$T_{\m\n}$ 
  and number density current conservation $J_\mu=n\;u_\m $
  generate the dynamical equation of motion of the fluid
\be
\nabla^\nu T_{\mu\nu}=0 \, ,\qquad \nabla^\mu J_\mu=0 \, .
\ee
 with $n$ representing the number particle density.
Indeed projecting the EMT conservation equations along  and 
   orthogonal to $u^\mu$ %direction %to $u^\mu$ 
%by using the projector  $h_\mu^\nu=\delta_\mu^\nu+u_\mu\,u^\nu$,
we have 
\bea\label{eqpf}
\dot \rho+\theta\;(p+\rho)=0,\qquad D_\mu\,p=-(p+\rho) \;{\cal A}_\mu\;
\eea
that with eq (\ref{trt}) constitute the equation of motion of the fluid.
{ To stress the interconnection in between the field formulation and the fluid formulation
is the fact that the equation of motion of the  scalar fields are given exactly by the conservation laws of the EMT $T_{\mu\nu}$ and the conserved current $J_\mu$ as it happens in hydrodynamics. 
Hydrodynamical constitutive eqs that give the structure of $T_{\mu\nu}$ and  $J_\mu$  as a function of energy, pressure, entropy, etc. are, in the field theory approaches, condensed inside the function ${\bf U}$. All of this is similar to the Lagrangian approach of classical fields where the eqs of motion for many variables can be compressed inside a single functional, the Lagrangian. }
\\
The thermodynamical dictionary that relates composite operators to thermodynamical quantities
was already studied in ref.\cite{Ballesteros:2016kdx}
%for a Perfect Fluid described by the potential $U(b,\,Y)$ is given by
 \bea\label{id1}
 {\rm pressure} && p={\bf U}-b\;{\bf U}_b,\qquad {\rm energy \;density} \;\;\rho=-{\bf U}+Y\;{\bf U}_Y\\
\label{id2}{\rm density} && n=n_0\;b,\qquad {\rm chemical \;potential} \;\;\mu=-\frac{{\bf U}_b}{n_0}\;\\
\label{id3}{\rm Temperature} && T=T_0\;Y,\qquad {\rm entropy \;density} \;\;s=\frac{{\bf U}_Y}{T_0}
 \eea
 where $n_0$ and $T_0$ are constants normalising factors.
 Finally we can identify the potential $U$ to be proportional to the Free energy $F$
 \be
 F=\rho-T\;s=-{\bf U}
 \ee
 Euler relation, First law of thermodynamics and Gibbs-Duhem relation are given by
 \be
 p+\rho=T\;s+n\;\mu,\quad d\rho=T\;d s+\mu\;dn,\quad dp=s\;dT+n\;d\mu
 \ee
 and are exactly satisfied with our identifications (\ref{id1},\ref{id2},\ref{id3}).
The differential structure of the pressure was  studied in  ref.\cite{Celoria:2017xos} and is given by
\bea\label{dp}
 dp  &=&
 \left.\frac{\partial p}{\partial \rho}\right|_{\sigma} \;d
\rho+\left.\frac{\partial p}{\partial \sigma}\right|_{\rho}\; d \sigma\;\equiv\;
 c_s^2\; d \rho+ c_\rho^2\;d\sigma
\, ;
%\\
\eea
 (we also define $ \mathbf{\Gamma}\equiv  c_\rho^2\;d\sigma$)
%
%c_s^2= \left.\frac{\partial p}{\partial \rho}\right|_{\sigma},\quad
 %c_\rho^2=\left.\frac{\partial p}{\partial \sigma}\right|_{\rho}
and  factor out an adiabatic contribution, proportional to $\delta\rho$  with $c_s^2$   the adiabatic sound speed
 and the non adiabatic contribution $\mathbf{\Gamma}
$ ~\cite{Kodama:1985bj} can be further factorised for time and space derivatives as 
\bea
 && \dot p
 =  c_s^2 \;\dot \rho,\qquad D_\a\, p=c_s^2\;D_\a \rho +
c_\rho^2\, D_\a\, \sigma \, ;
\eea
specifically we have (for $\rho+p\neq 0$)%where $c_s^2$ is the sound velocity 
%where in the second \ref{ch:stab}
\bea
c_s^2  =    %\frac{(U_Y-b\;U_{bY})^2-b^2\;U_{b^2}\;U_{Y^2}}{{  U_{Y^2}}\;(\rho+p)}=
\frac{c_b^4\;Y^2\;{\bf U}_{Y^2}-b^2\;{\bf U}_{b^2}}{\rho+p},
%=\frac{\a^2\; T+c_V\;\k_T}{c_V\;\k_T^2\;(\rho+p)}=\frac{c_p}{c_V\;\k_T\;(\rho+p)},
 \qquad
%\qquad % \label{cs}
%\ee
%\be
%&& 
c_\rho^2\; =  b\;Y\;(c_b^2-c_s^2),\qquad 
c_b^2 = \frac{{\bf U}_Y-b\;{\bf U}_{bY}}{Y\;{\bf U}_{Y^2}} \, . 
\label{cb}
\eea
{  A final important parameter (mainly to estimate the cosmological impact of the system on the Universe evolution) is the effective equation of state  given by the ratio
\be
w=\frac{p}{\rho}
\ee
Perfect fluids with $w=const$   are generate by the following potentials:
\be\label{eqw}
{\bf U}=b^{1+w}\; U\!\!\left(b^{-w}\,Y\right),\qquad w={\rm const}
\ee with $U$ a generic function.
We see that  a DM fluid with $w=0$ is generated by a potential of the form ${\bf U}=b^{}\; U(Y)$ while for a DE fluid with $w=-1$ we have ${\bf U}= U(b\;Y)$.
}
 Once the perfect fluid potential is given ${\bf U}(b,Y)$, one can
compute $c_s^2$, $c_b^2$, $c_\rho^2$ and $w$,    then $\mathbf{\Gamma}
$ is known in a fully non
perturbative way.  
In appendix (\ref{ch:stab}) we revisit the thermodynamical stability conditions applied to such a general formalism. 
\\
{  Stability analysis for the perturbations of a media %for short wave lengths 
can be developed  around a Minkowski space where the pressure and the density of the system are constant in space and time.
In fact analysing also the perturbations around a FLRW background in the short wave length limit,
% it is possible to see that 
 the same stability conditions are recovered  for a medium on Minkowski spacetime \cite{Celoria:2017hfd}.
%Such analysis  was carried out for example in \cite{Celoria:2017hfd} where 
We   consider the goldstone perturbations $\pi^A$ %around Minkowski space 
where
$\Phi^A=x^A+\pi_A$ (A=0,...,4)
(we can further decompose the spatial components as $\pi_a=\partial_a \pi_L+V^a$ with 
$\partial_a  V^a=0$, $a=1,2,3$) so that  we have two scalar dof  ($\pi_0$ and $\pi_L$)  and two divergence-less vector dof $V^a$.
In general the goldstone quadratic Lagrangian can be parametrised using   four mass parameters $\hat M_i %=M_{Pl}^2 \, M_i
$, $i=0,1,2,3,4$  defined in terms
of the derivatives of the medium Lagrangian (\ref{actS}), their explicitly
expression  is given in appendix of ref.\cite{Celoria:2017hfd}.
From the  mechanical point of view we have: perfect fluids when $\hat M_{1,2}=0$,
  solids for $\hat M_1=0$, superfluids for $\hat M_2=0$ and supersolids when
  $\hat M_{1,2}\neq0$. So,  in our case,    a perfect fluid gives the following quadratic Lagrangian \footnote{The explicit expressions in such a case are $\hat M_0=\frac{Y^2}{2} U_{Y^2}$, $\hat M_3=\frac{b^2}{2} U_{b^2}$, $\hat M_4=\frac{b\,Y}{2} U_{bY }-\frac{ Y}{2} U_{Y }$ evaluated on Minkowski space where $b=Y=1$}
\bea
L^{(2)} &=&
% \frac{k^2}{2} \left( \hat{M}_1 +p+ \rho\right)\pi_l'{}^2 +
% \hat{M}_0 \,  \pi_0'{}^2 + \frac{k^2}{2} \, \hat{M}_1
%\pi_0^2 + k^2 \left(\hat{M}_1-2 \, \hat{M}_4\right) \pi_0' \, \pi_l 
%+ k^4 \left(\hat{M}_3-\hat{M}_2\right) \pi_l^2 \, 
%%=\\&&
 \left(\frac{(\bar\rho+\bar p)}{2\,M_{Pl}^2} \,\left((\vec\nabla\dot\pi_L{})^2 +\vec V^2\right)+
   \hat{M}_0\,  (\dot \pi_0{})^2  +2\,\hat{M}_4\;\left(  \dot \pi_0 \;\nabla^2 \pi_L \right)
+\hat{M}_3     (\nabla^2\pi_L)^2\right)
\eea
The dispersion relation for transverse modes $V^a$ is $\omega_V(k) =0$
while for the two scalar dof (note that in the scalar sector this lagrangian has the same structure of a gyroscopic system see \cite{giro}) we have $\omega_0(k) =0$ and 
$\omega_L(k)=\bar c_s\;k$
where $\bar c_s^2= \frac{2\,(  \hat{M}_4^2-
  \hat{M}_0 \,  \hat{M}_3 )}{ \hat{M}_0 \, 
  (\bar p+\bar \rho )}$ is the sound speed.
  So, in general, only longitudinal waves  propagates with a finite sound speed. 
   The transverse  fluctuations correspond to an infinitesimal version of a fluid vortex.
 The fact that, in the EFT approach to perfect fluids,
    some field excitations lack of gradient energy
   results  an open problem \cite{rat}, in particular once we try to quantise the system
   and it results strictly related to the volume diff symmetries of the world space $\Phi^a$.
  % It can be shown that the \pi_0 field never propagate being the potential a function of only $\dot \pi_0$
  \\
 For scalar perturbation the bridge in between  the $\pi_{0,L}$ fields and the fluid approaches with  observables like
the energy density perturbation $\delta \rho$, the
pressure perturbation $\delta p$, the entropy per
particle perturbation $\delta \sigma$, or the velocity $\vec v$
is given by the following   Fourier transformed relationships  ($x=\bar x +\delta x$ and $x=\rho, p, \sigma$ with  $\bar x$   constants).
\bea
%\begin{split}
\!\!\!\!\! &&\delta \sigma = -2 \;M_4 \left(\hat c_b^2\;\dot \pi_0{} + k^2 \, \pi_L \right), %=2 \; \hat{M}\;\left(\dot\pi_0{} -k^2 \, \pi_l \right)\, ,
    \;\;% \\&& 
     \delta\rho = \delta \sigma - (\bar \rho+\bar p)\, k^2 \, \pi_L ,
\\&&
\delta p = \bar c_s^2 \, \delta \rho + (\bar c_b^2 -\bar c_s^2) \, \delta \sigma,
%\end{split}
\;\;\;\;\;\;\;\;\;
\vec v=-\dot{\vec V}-\vec{\nabla}\dot \pi_L
\eea
with $\bar c_b^2=-  \hat M_0/\hat M_4 $.% - \frac{\hat M_0}{\hat M_4}$ 
The equations of motion for the  density and  the entropy  per particle fluctuations describe the evolution of the system
\bea
\delta\ddot\rho+\bar c_s^2\;k^2\;\delta\rho+(\bar c_s^2-\bar c_b^2)\;\delta\sigma=0,\qquad \delta\dot \sigma=0
\eea
Unacceptable exponential instabilities are present  %at the border 
for $\bar c_s^2<0$.
While for $\bar c_s^2\geq 0$ we have the following solutions   
\bea
&& \delta \sigma=const
\\
&&{\rm for}\;\;\bar c_s^2>0\to \delta\rho=\bar c_1\; \cos(\bar c_s\;k\; t)+
\bar c_2 \;\sin(\bar c_s\;k \;t)+\left(1-\frac{\bar c_b^2}{\bar c_s^2}\right)\;\delta\sigma
\\
&&{\rm for}\;\;\bar c_s^2= 0\to\delta\rho=\bar c_1 +\bar c_2 \;t - \bar c_b^2\;k^2\;t^2\;\delta\sigma
%\\
%&&{\rm for}\;\;\bar c_s^2< 0\to\delta\rho= 
\eea
with $\bar c_{1,2}$ the initial conditions for the energy fluctuations. 
%Unacceptable exponential instabilities start %at the border 
%for $\bar c_s^2<0$.
Note that for zero sound speed   $\bar c_s^2= 0$  there are power law growing mode in time.
% in presence of entropic modes.
%In such a case the validity of the perturbation theory is frozen to the interval $k\;t\leq 1$.
We stress such a  specific case because the system we are interested in is characterised exactly  by the configuration with $\bar c_s^2=0$ and $\bar c_b^2=-1$, see next chapter.
% The energy 
% \be
% E=\hat M\left(\dot \pi_0^2-k^4 \,\pi_L^2 \right)+\frac{\bar \rho+\bar p}{2}\left( (  
% \dot{\vec{V}})^2+\dot \pi_L^2\right)
% \ee

}

   \section{The Next to Minimal $\Lambda$CDM Model (N$\Lambda$CDM )}
  \label{ch:NLCDM}

In order to describe the Matter-Dark energy dominated period of the Universe evolution, we introduce
 a   single 
 Perfect fluid  where both DM and  DE take place.
{ Thanks to the structure of potential with $w$ constant in (\ref{eqw})  we can add two term, one corresponding to $w=-1$ (DE) and one to $w=0$ (DM) to get a system describing
 DE and DM all together.  }
 The dynamics of such a system is described by the following action, already studied in ref.\cite{Celoria:2017xos}
   \be\label{lagU}
   S=\int d^4x\;\sqrt{g}\;\left(U(b\;Y)-m\;b \right)
   \ee
   where   $m$ is  a mass parameter multiplying the density field $b$ (we will use the notation $m\,b =n$) and the potential $U(b\,Y)$  corresponds to a $\Lambda$ perfect fluid (whose details are given in appendix (\ref{ch:CC})).
     In Minkowski space the perturbations of such a system are characterised by $\hat M_{0,3,4}=\frac{b \,Y }{2}U(b \,Y )$ (where $b =Y =1$) that gives $\bar c_b^2=-1$ and $\bar c_s^2=0$. 
     So, no dof is propagating being $\omega_{V,0,L}(k)=0$ and no ghost like pathologies are present.
   %   The function $U$ is a special potential that generates equation of motion   identical to a   CC,  providing, when alone, an exact de Sitter space time.
 On a generic space time background   the corresponding EMT is
   \be
   T_{\m\n}=\underbrace{(U-b\,Y\;U')}_{CC}\;g_{\mu\nu}+\!\!\!\!\underbrace{n}_{Matter}\!\!\!\!\;u_\mu\;u_\nu
   \ee
   where the CC ($w=-1$) and the Matter ($w=0$) content is explicitly shown.
      Energy density,  pressure and  thermodynamical parameters are given by
       \bea\label{par}
&& \rho=\Lambda+n,\quad p=(U-b\,Y\;U')=-\Lambda\\
&& s=b\;U',\quad \sigma=U',\quad\mu=m-Y\;U'   \\
&& c_s^2=0,\quad c_b^2=-1
   \eea
   that implies (with $Z\equiv b\,Y$)
   \bea
&&  d\sigma=U''\;dZ ,\qquad dp= -Z\;U''\;dZ\;\;\Rightarrow\;\;dp=-Z\;d\sigma\\
   &i.e.&\qquad \dot p=-b\,Y\; \dot \sigma \quad {\rm and}\quad D_\mu p=-Z\; D_\mu \sigma
   \eea
   The EMT conservation is providing the exact eqs of motion \footnote{Note that in the inhomogeneous vacuum energy models of \cite{Wands:2012vg} they study   the cases where $\dot \Lambda\neq 0$ on a FRW background.}
   \bea\label{as}
       \nabla_\m\Lambda= n\;{\cal A}_\m 
   %-Z\;U''\;\nabla_\mu Z &+& n\;{\cal A}_\mu=0
   \;\;\Rightarrow\;\;
   \left\{ 
\begin{array}{l }
\dot \Lambda=0 \to \dot \sigma=0\\
   {\cal A}_\mu= \frac{ D_\mu \Lambda}{n}\\
\end{array}%\bigg \}
\right.
   \eea
%   where we used the density conservation eq (that olds off shell)
%   \be
%   \dot b +\theta\;b=0
%   \ee
where the first eq. results from the projection along the $u^\m$ direction  ($u^\m \nabla_\m\Lambda\equiv \dot\Lambda$ and  
 $u^\m  {\cal A}_\mu=0$, ${\cal A}_\mu$ being the acceleration field (\ref{gradu})), the second eq. instead is given by the orthogonal projection obtained multiplying by   $h_\n^\m=\delta_\n^\m+u_\n\,u^\m$ and $D_\n\equiv h_\n^\m \nabla_\m$ (see also Appendix A).
   If we want to describe the present DM/DE transition in a FRW framework, we have just to tune the parameter to the present energy abundance
   %and take a zero value  for the vector fields on the background, so that $\Lambda$ is a true CC (as first approximation)
   \be
  n_0 =3\;H_0^2\; \Omega_m,\quad   \Lambda =3\;H_0^2\; \Omega_\Lambda
   \ee
   where $H_0$ is  the present Hubble constant, $\Omega_m$ and $\Omega_\Lambda$ the fraction of DM and DE at present time ($\Omega_m+\Omega_\Lambda=1$).
   So that
 \bea\label{hh}
  %\theta=3\;H,\;\;
  \hspace{-0.5cm} \rho=3\;H_0^2\;\left(\frac{\Omega_m}{a^3}+\Omega_\Lambda\right) ,\;\;p=-3\;H_0^2\;  \Omega_\Lambda ,\;\;n=\frac{n_0}{a^3} 
,\quad
     w= -\frac{ \Omega_\Lambda\;a^3 }{ \Omega_m +\Omega_\Lambda\;a^3},\quad c_s^2 =0
     \eea
     and it corresponds exactly to the $\Lambda$CDM predictions for the last period of $DM/DE$ domination.
     Now that we set the background behaviour we can look to the structure of the perturbations.
     In ref.\cite{Celoria:2017xos} it was shown that, for a FRW background, the comoving curvature perturbation ${\cal R}$  at large scales grows up as $a^3$ at late times (\ref{Rc}).
     To show directly the instability of the FRW background we can write the full evolution equation
         for a specific  gauge invariant operator: the acceleration vector ${\cal A}_\mu$.
  The time evolution of the acceleration ${\cal A}_\mu$, for a generic perfect fluid is given by the eq. (see {  appendix A})\footnote{We used the formula
\bea 
\dot{  (D_\mu \sigma)}=
   \;u_\m\; ({\cal A}^\a\;D_\a \sigma) -\frac{\theta}{3}\;D_\m\sigma-{  \sigma}_{\m\a}\;D^\a\sigma+
   \omega_{\m\a} \;D^a\sigma
  \eea}
   \bea\label{eqa}
\dot {\cal A}_\m-\left(\frac{2}{3}+2\;c_s^2\right) \;\theta\;{\cal A}_\m-D_\mu(\theta\;c_s^2)={\cal A}^2\;u_\m+(\omega_{\m\a} -\sigma_{\m\a})\;{\cal A}^\a
\eea
Note that the left handed side of eq.(\ref{eqa}) is first order in perturbations around FRW while the right handed side is at least second order.
Then for the N$\Lambda$CDM fluid   $c_s^2=0$  (\ref{par})   and
\be
\dot{\bf A}- \frac{2}{3}  \;\theta\;{\bf A}=  - \;\frac{{\cal A}^\a\;\sigma_{\a\b}\;{\cal A}^\b}{\bf A},\qquad {\bf A}^2\equiv
{\cal A}_\a\;{\cal A}^\a
\ee
     that at first order gives  $ {\bf A} \propto a^{2}$ so that ${\cal A}_\m\propto a^3\sim \frac{1}{n}$,
i.e. the acceleration   growths as the third power of the scale factor, at any scales and in any DM-DE dominated period, entering soon or later in a non perturbative regime and destabilising the FRW background.
  As discussed in ref.\cite{krasinski},   one of the necessary conditions for a FRW limit  there is  a zero acceleration.\footnote{
 Following \cite{Celoria:2017xos},   the Master equation for the Bardeen 
 potential $\Phi=\Phi(a)$  on   FRW (\ref{hh}) is
  \bea
  \Phi ''+\frac{
   \left(10\; a^3\; \Omega
   _{\Lambda }+7\; \Omega
   _m\right)}{2 \;a\;(\;a^3 \;\Omega
   _{\Lambda }+ \Omega
   _m)}\;\Phi '+\frac{3\;
   a   \;\Omega _{\Lambda
   }}{a^3 \;\Omega _{\Lambda
   }+\Omega _m}\;\Phi=\frac{3\; a\; \sigma }{2 \;H_0^2\;
   \left(a^3 \;\Omega _{\Lambda
   }+\Omega _m\right)}
  \eea
   ($\Phi'=\partial\Phi/\partial a$ etc.) whose non homogenous solution   in    Matter and in   $\Lambda$  dominated  periods behaves as
  \bea
  \Phi|_{M}\propto \Phi_0\; \frac{a^3  }{\Omega _m},\qquad 
  \Phi|_{\Lambda}\propto \frac{  \Phi_0}{\Omega_{\Lambda}}
  ,\qquad 
   \Phi_0=\frac{ \sigma }{ H_0^2 }
  \eea
  %that shows   a cubic (with the scale factor $a$)  growing perturbation.
  %Note that, being $c_s^2=0$, the only scale dependence for the Bardeen is, for the homogeneous eq 
  While for the comoving curvature perturbation ${\cal R}=-\Phi-H\;v,\;v=\frac{2\;(a\;\Phi'+\Phi)}{3\;(1+w)}$ we get
 \bea
 {\cal R}|_{M}\propto  \Phi_0\;\frac{ \Omega_{\Lambda }}{\Omega_m}\;a^6,\qquad 
   {\cal R}|_{\Lambda}\propto  \Phi_0\;
   \frac{ \Omega_{\Lambda }}{\Omega_m}\;a^3\label{Rc}
\eea
so that at late time, $a\to \infty$ during the DE domination era, also if the Bardeen potential remains constant, the comoving scalar  curvature (due to the presence of the scalar vector component $v$) grows as $a^3$ (see also (\ref{eqa})), soon overwhelming the perturbative limit. 
}
   \section{1+1+2 formalism}     \label{ch:112}
      
   FRW background is a space geometrically isotropic about all the fundamental world lines implying zero shear, vorticity and acceleration.
A non zero value for   these quantities would pick out preferred directions in the 3-$d$ space orthogonal to the vector $u^\a$.
   To describe relativistic cosmology around a FRW the use of the covariant 1+3 approach (see \cite{Ellis:1998ct})  results quite   a powerful tool,   especially for the definition  of the gauge-invariant and covariant perturbation formalism. 
%   This perturbation formalism works extremely well in cosmological applications when the background model is homogeneous and isotropic, that is of Friedman-Lema?itre-Robertson-Walker (FLRW) type.
  The space time is     splitted in {\it time} and {\it space} relative to the fundamental observer represented by
    the timelike unit vector field $u^\a$ ($u^2=u^\a\;u_\a = -1$), representing the observer's 4-velocity.
    In this way the covariant 1+3 threading irreducibly splits any 4-vector into a scalar part parallel to $u^\a$ and a 3-vector part orthogonal to $u^\a$.
    Furthermore, any second rank tensor is covariantly and irreducibly split into scalar, 3-vector and projected symmetric trace-free  3-tensor parts.
  The previous FRW analysis  showed the presence of   
  a gauge invariant vector field ${\cal A}_\m$, the acceleration,  that growth very fast in time at all scales.
  In this section we change the background hypothesis going from an homogeneous and isotropic model
  (FRW) with a preferred time-like vector field $u^\a$
  to a  {\it Locally Rotationally Symmetric} (LRS) spacetime where, in addition to the time-like vector field $u^\a$, it exists also a  covariantly defined unique preferred spatial direction, $v^\a$, that in our case 
  is the direction of the acceleration field ${\cal A}^\a$.
  To describe  the structure of such a kind of spaces the 1+1+2 formalism, developed in \cite{vanElst:1995eg} (see also \cite{Clarkson} for developments),  is therefore ideally suited for a covariant description   in terms of invariant scalar quantities that have physical or direct geometrical meaning.
The preferred spatial direction in the LRS spacetimes constitutes a local axis of symmetry and    is just a vector pointing along the axis of symmetry and is thus called a  {\it radial  vector}. 
  Since LRS spacetimes are defined to be isotropic, this allows for the vanishing of all 1+1+2 vectors and tensors, such that there are no preferred directions in the sheet (the 2-$d$ space orthogonal both to   $u^\a$ and $v^\a$). 
  Thus, all the non-zero 1+1+2 variables are covariantly defined scalars.
  The variables needed to describe the LRS space form an irreducible 1+1+2 set 
  (see {\it Appendix B} for the mathematical details):
\begin{itemize}
\item From a generic EMT we have 
  density $\rho$, pressure $p$, the projected energy flow ${\cal Q}$ and the projected anisotropic scalar $\Pi$ variable. 
  
 \item The split of the (1+3) kinematical variables (related to the gradient of $\nabla_\a\,u_\b$) gives the projected
 acceleration ${\cal A}$, shear $\Sigma$ and vorticity $\Omega$. 
  \item The decomposition of the gradient of  $\nabla_\a\,v_\b$ gives the structure of the 2-$d$ sheet space 
   with the sheet expansion $\phi$ and the twist $\xi$.
   
   \item Finally the various projections of the Weyl tensor  generate two scalars: the electric ${\cal E}$ and the magnetic ${\cal H}$ one.
  
  \end{itemize}
 So, the geometrical scalar variables that fully describe LRS spacetimes are
\be\label{subD}
{\cal D}_2=\{\;\underbrace{\rho,\;p,\;\Pi,\;{\cal Q}}_{EMT},\;
\underbrace{{\cal A},\;\theta,\;\Sigma,\;\Omega}_{\nabla u},\;\underbrace{\phi,\;\zeta}_{\nabla v}, \;\underbrace{{\cal E},\;{\cal H}}_{Weyl}
\}
\ee
  A subclass of the LRS spacetimes, called LRS-II, contains all the LRS spacetimes that are rotation free. As consequence in LRS-II spacetimes the variables $\Omega, \;\zeta$ and ${\cal H}$ are identically zero and the rest of the variables in ${\cal D}_2$
%$\{\rho,\;p,\;\theta,\;{\cal Q},\;\Pi,\;{\cal A},\;\Sigma,\;{\cal E}\}$
fully characterise the kinematics. 
\vspace{0.2cm}
\\
The   \underline{LRS\; spaces\; for\; Perfect\; Fluids} (${\cal Q}=\Pi=0$) can be further divided into three classes
\begin{itemize}
\item {\bf LRS Class I}:
$\boxed{\Omega\neq0}$ with $\xi=\theta=\Sigma=0$ and for any scalar quantity $\dot f=0$.
These model are stationary, non expanding, not distorting, shear free.
\item {\bf LRS Class II}: $\boxed{\Omega=\xi=0}$ then also the    magnetic part  of the Weyl curvature tensor  ${\cal H}$ is 
vanishing. These models contain spherical, hyper-spherical and plane symmetric (cylindrical) solutions. 
\item {\bf LRS Class III}: $\boxed{\xi\neq0}$ with $\Omega=\phi={\cal A}=0$ and for any scalar quantity $\hat f=0$.
These model are spatially homogeneous.

\end{itemize}

Singh et al. \cite{Singh:2016qmr} found a new class of LRS spacetimes that in presence of non zero heat flux have nonvanishing rotation and spatial twist.
%The most general rotating and twisting LRS spacetimes was found by Singh et al \cite{} which allows for a detailed analysis for gravitational collapse. 
%Explicit forms of the LRS spacetime metrics and self-similar vectors were identified by Van den Bergh \cite{}.
  \begin{table}[htp]
\begin{center}
\begin{tabular}{|c||c|c| c| c|}
 \hline
          & Matter& Geometry:$\;\nabla\;u$ & $\nabla v$ & Weyl\\   \hline \hline
LRS &   $\rho,\quad p,\quad \Pi,\quad {\cal Q},\quad n$ & $\theta ,\quad {\cal A},\quad\Sigma,\quad \Omega$ & $ \phi,\quad \xi$ & $
{\cal E},\quad {\cal H} $
\\ \hline
LRS II  PF&   $\rho,\quad p,\quad n%,\quad \Pi,\quad {\cal Q}
$ & $\theta ,\quad {\cal A},\quad\Sigma%,\quad \Omega
$ & $ \phi %,\quad \zeta
$ & $
{\cal E}%,\quad {\cal H} 
$\\ \hline
LRS II  PF Stationary&   $\rho,\quad p,\quad n%,\quad \Pi,\quad {\cal Q}
$ & $%\theta ,\quad 
{\cal A}%,\quad\Sigma%,\quad \Omega
$ & $ \phi %,\quad \zeta
$ & $
{\cal E}%,\quad {\cal H} 
$\\ \hline
$KS_{\kappa=1} $&  $\rho,\quad p,\quad n$ & $\theta,\;\Sigma $ & &  $
{\cal E}%,\quad {\cal H} 
$\\
\hline
$LTB_{\kappa=1}$&  $\rho, \quad n$ & $\theta,\;\Sigma $ & $\phi$ &$
{\cal E}%,\quad {\cal H} 
$  \\
\hline
$FRW_{\kappa=1}$&  $\rho,\quad p,\quad n$ & $\theta $ &$\phi$ &  \\
\hline
\end{tabular}
\caption{Background parameters needed to describe various LRS space times. 
For the column "Matter" we have the non zero background scalars   relative to the EMT and the particle density current. Into the column "Geometry: $\nabla u$" we have the non zero background kinematic form factors relative to the gradient of the time like  $u^\a$ vector.
Into the column "$\nabla v$" we have the non zero background kinematic form factors relative to the gradient of the space like $v^\a$ vector.
Into the column "Weyl" we have the non zero background form factors relative to the Weyl tensor.
KS=Kantowski-Sachs models,   LTB=Lema$\hat{\rm i}$tre Tolman Bondi models, FRW=Friedmann Robertson Walker models. The $\kappa$ factor is related to the topology of the sheet 2-$d$ space.
}
\end{center}
\label{tab1}
\end{table}%
\\
For the analysis of our model we choose the   space like unit vector $v^\a$   as the spatial direction of the acceleration field:
 \footnote{Here some useful identities
  \bea 
  \dot {\cal A}_\a&=&\dot {\cal A}\;v_\a+{\cal A}\;\dot v_\a,\quad
  v ^\a\dot v_\a = 0,\quad u^\a\dot v_\a = -{\cal A} ^\a  v_\a    \\
  && u^\a \dot {\cal A}_\a={\cal A}\;u^\a \dot v_\a=-{\cal A}\;{\cal A}^\a  v_\a=-{\cal A}\;\dot {\cal A}\\
  && \;\;\dot {\cal A}=v ^\a   \dot {\cal A}_\a,
   \quad {\cal A}_\a\dot {\cal A}^\a={\cal A}\;\dot{\cal A}
\eea
}
  \bea  \label{accv}
   {\cal A}_\a \equiv {\cal A}\;v_\a,  \quad v^2=1,\quad u^\a\;v_\a=0
  \eea
  Due to the presence of two special vectors, one time-like $u^\a$ and the other space-like $v^\a$ we have two  {\it directional derivatives} for a generic scalar function $f$
  \bea\label{dirD}
  \dot f \equiv u^\a\;\nabla_\a f ,\qquad \hat f \equiv v^\a\;\nabla_\a f,  
  \eea
The EMT conservation eqs for a PF and the differential structure of the pressure (\ref{dp})  give
\bea 
&& dp=c_s^2\;d\rho+c_\rho^2\;d\sigma,
\\
&&\dot\rho=-\theta\;(\rho+p) ,\quad
 \dot p=c_s^2\;\dot\rho,\quad\
\dot\sigma=0,\quad \dot n+\theta\;n=0,\quad \dot T-\theta\;T=0\\
&&
\hat p=-{\cal A}\;(\rho+p) ,\quad \hat p=c_s^2\;\hat\rho+c_\rho^2\;\hat\sigma \eea
 The mixed    derivatives $\hat{\dot f}$  or  $\dot{\hat f}$ for the density and pressure reads (see eq.(\ref{mix}) and app. (\ref{mixed}))
 \bea
% {\rm mixed} \; \; \;\hat{} \;-\; \dot{} \; \; \;{\rm derivatives\; in\; LRSII }\\
&&\hat{\dot\rho}=-\hat\theta\;(\rho+p)-\theta\;\left(\hat \rho- {\cal A}\;(\rho+p)\right)
%,\quad%\\&&
%\dot{\hat p}=-\left(\dot {\cal A}-{\cal A}\;\theta\;(1+c_s^2)\right)\;(\rho+p)
 ,\qquad
 % &&\dot{\hat p}=c_s^2\;\dot{\hat\rho}+c_\rho^2\;\dot{\hat\sigma}+\dot c_s^2\;\hat\rho+\dot c_\rho^2\;\hat\sigma
%,\quad
\hat{\dot p}=\hat{c_s}^2\;\dot\rho+c_s^2\;\hat{\dot\rho}
 \eea
Deriving eq. (\ref{dp})    we   get  (for $\rho+p\neq0$) the general expression for the evolution of ${ \cal A}$%cceleration
\footnote{Interestingly, for $c_s^2 \neq0$, we can rewrite eq.(\ref{dotA}) as function of %the entropy density 
$\sigma$ 
(replacing $\hat \rho$ with $\hat \sigma$)   
\bea
\dot{{\cal A}}&=&{\cal A} \;
\left( \left(\frac{2 
   }{3}+ 
   c_s^2\right)\;\theta-
   \frac{\theta \;
   c_s^2  \;(\theta +3  \;\Sigma )}{c_s^2 \;
   (\theta +3  \;\Sigma )-3 \;
   \dot{ c}^2_s}-\Sigma \right)+ \\&&
   c_s^2 \;
   \left( \frac{3  \;\theta ^2 \;
   \hat{c}^2_s}{c_s^2 \; (\theta +3  \;\Sigma
   )-3 \; \dot{c}^2_s}+\hat{\theta
   }\right)-
  \frac{\hat{ \sigma}\;
   \theta\;  c_s^2\; \left(c_{\rho }^2\;
   (\theta +3 \;\Sigma )-3\;
   \dot{c}^2_{\rho }\right)}{(p+\rho
   )\; \left(c_s^2\; (\theta +3\; \Sigma )-3\;
   \dot{c}^2_s\right)}-\theta \;
   \hat{c}^2_s
\eea
and in the limit of constant $c_s^2$ and $c_\rho^2$ (always $c_s^2\neq 0$) becomes
\bea
\dot{{\cal A}}&=&{\cal A} \;
\left( \left({ c}^2_s- \frac{1}{3}\right)\;  \theta   -  \Sigma    \right)+ 
   c_s^2 \;\hat{\theta}-
  \frac{\hat{ \sigma}\;
   \theta\;   c_{\rho }^2 }{(p+\rho
   ) } 
\eea
}
\be\label{dotA}
\dot {\cal A}=\left(\left(\frac{2\; }{3}+c_s^2\right)\;\theta-\Sigma+\frac{\dot c_\rho^2}{c_\rho^2}
\right)\;{\cal A}+c_s^2\;\hat \theta+
\left( \left(\theta+\frac{\dot c_\rho^2}{c_\rho^2}\right)\;c_s^2-\dot c_s^2
\right)\;\frac{\hat \rho}{p+\rho}
\ee

For the N$\Lambda$CDM  fluid we have  $c_s^2=0$ and $c_\rho^2=-1$ so that the acceleration evolves as
\be\label{eqA}\boxed{
\dot {\cal A} =
 \left (\frac{2\;\theta}{3} -\Sigma\right)\;{\cal A}  }
\ee
supplemented by the eqs (\ref{par},\ref{as})
\bea\label{pL}
p=-\Lambda,\qquad \dot \Lambda=0,\qquad \hat \Lambda=n\;{\cal A}
\eea
Note that a positive acceleration is related to a positive gradient of the cosmological constant (negative for the pressure).
The general evolution/propagation equations for the rest of the scalar functions can be classified    depending on the kind of covariant derivatives ( $\hat{} $ or $\dot{}$ )  and are obtained   from the Bianchi and Ricci identities respectively:
\begin{itemize}
\item {\it Evolution eqs}
\bea \label{eqe}
&&  \dot\Sigma=\frac{2}{3}\hat {\cal A}+\left(\frac{2}{3}\;{\cal A} -\frac{\phi}{3}\right)\,{\cal A}-
\left(\frac{2}{3}\;\theta+\frac{1}{2}\;\Sigma\right)\,\Sigma-{\cal E}\\\nonumber
 &&\dot \phi=\left(\frac{2\;\theta}{3} -\Sigma \right)\;\left({\cal A}-\frac{\phi}{2}\right)
 \\&& \dot{\cal E}=\left(\frac{3\;\Sigma}{2} -\theta \right)\;{\cal E}-\frac{\rho+p}{2}\;\Sigma\nonumber
\eea
\item {\it Propagation eqs}
\bea \label{eqp}
  &&\hat\Sigma-\frac{2\;\hat\theta}{3} +\frac{3\;\phi}{2} \;\Sigma=0\\\nonumber
  &&\hat\phi=-\frac{ \phi^2}{2} +\left(\frac{\theta}{3} +\Sigma\right)\;
  \left(\frac{2\;\theta}{3} -\Sigma\right)-\frac{2\;\rho}{3} -{\cal E}\\\nonumber
  &&\hat{\cal E}-\frac{\hat\rho}{3}=-\frac{3}{2}\;\phi\;{\cal E}
  \eea
\item{\it Propagation/Evolution eqs}
\bea\label{eqpe}
&&\dot \theta=\hat {\cal A}+\left({\cal A}+\phi \right)\;{\cal A}-\frac{1}{3}\;\theta^2-\frac{3}{2}\;\Sigma^2-\frac{1}{2}\;(3\;p+\rho) 
\eea

%\item{Constraint eq}

\end{itemize}   
  Being the vorticity zero, we have that the vector $u^\a$ is hypersurphace orthogonal to space like 3-$d$ surfaces with 3-curvature $^{(3)}R$ given by
  \be
  \;^{(3)}R=-2\left(\frac{\hat \phi}{2}+\frac{3}{4}\;\phi^2-{\cal K}\right)=2\;\left(
  \rho-\frac{\theta^2}{3}+\frac{3}{4}\;\Sigma^2
   \right)
  \ee
  with ${\cal K}$ the Gaussian curvature of the 2-$d$ sheet given by
  \be
  {\cal K}=\frac{\rho}{3}-{\cal E}+\frac{\phi^2}{4}-\frac{1}{4}\;\left( \Sigma-\frac{2}{3}\;\theta\right)^2
  \ee
  characterised by the following evolution/propagation equations
  \be\label{eqk}
  \hat {\cal K}=-\phi\;{\cal K},\qquad \dot {\cal K}=\left( \Sigma-\frac{2}{3}\;\theta\right)\;{\cal K}
  \ee
  The ${\cal K}$ variable and his evolution eqs (\ref{eqk}) can be used to replace one of the variable in  
  subset ${\cal D}_2$, (\ref{subD}).
   It is interesting to rewrite a subset of coupled evolution equations using the $S\equiv \Sigma- \frac{2}{3}\;\theta$ variable:
 \bea\label{eqSS}
&& % \!\!\!\!\!  \!\!\!\!\! 
  \dot {\cal A}=-S\;{\cal A},\qquad \dot {\cal K}=S\;{\cal K},\qquad \dot p=0,\quad
  \\&&\nonumber
\dot\phi=\frac{S}{2}\;\phi-S\;{\cal A}, \quad
  \dot S=p+{\cal K}+\frac{3}{4}\;S^2-\phi\;{\cal A}-\frac{\phi^2}{4} 
  \eea
  Rearranging the above eqs we can get
  \be
  \dot{({\cal K}\;{\cal A})}=0,\qquad \dot \phi-\dot{\cal A}=-\frac{\dot{\cal A}}{2\;{\cal A}}\;\phi
  \ee
  whose non perturbative solutions, as a function of two boundaries $f_{0,1}$ functions, are
  \footnote{Strictly speaking we have to distinguish the positive/negative acceleration cases with the corresponding solutions
   $\phi= \frac{2}{3}\;{\cal A}+\frac{\mathit{f}_0}{\sqrt{\pm\cal A}}$. In order  
  to have compact notations we wrote only  $\sqrt{ \cal A}$ assuming that
    $ \mathit{f}_0 $ will be pure complex when  ${\cal A}<0$.}
  \bea\label{kf}
  \boxed{{\cal K}=\frac{\mathit{f}_1}{{\cal A}} ,\qquad \phi= \frac{2}{3}\;{\cal A}+\frac{\mathit{f}_0}{\sqrt{\cal A}}},\qquad \dot{\mathit{f}}_{0,1}=0\eea
  
 As soon as $f_{0,1}\neq0$  we can rewrite the full system of eqs (\ref{eqSS}) as a single evolution equation for ${\cal A}$\footnote{The fact that such eq. is higher order in time doesn't means that the theory is of higher order (i.e. affected by ghost like pathologies) it is simply the fact that we integrated out many fields from the system of first order eqs in  (\ref{eqSS}).}:
  \bea
 \dddot{\!\!\!\mathcal{A}}-
   \frac{\dot{\cal A}\;\ddot{\mathcal{A}} }{2\;{\mathcal{A}}} +
   \frac{1}{4}\;
   \frac{\dot{\mathcal{A}}^3}{  \mathcal{A}^2}
   +\dot{\cal A}\;\left( p-2 \; \mathit{f}_0\;\sqrt{\mathcal{A}}-\frac{7}{3}\;
   \mathcal{A}^2\right)=0  \eea
   and using   
  $ %
  % \hat{\phi={\cal K}-\frac{3}{4}\;S^2-S\;\theta-\rho-\frac{3}{4}\;\phi^2,\qquad
   \hat{\cal K}=-\phi\;{\cal K} 
   $, we can write the eq. for the spatial gradient of ${\cal A}$ as
   \bea
    \boxed{  \hat{\cal A} =\frac{2}{3}\;{\cal A}^2+f_0\;\sqrt{\cal A}+\frac{\hat{f}_1}{f_1}\;{\cal A}}
   \eea
   while the eqs for the other variables result 
   \bea
   \dot \theta&=&-\theta^2-2\;\;S\;\theta-\frac{3}{2}\;S^2-\frac{3\;p+\rho}{2} +2\;f_0\;\sqrt{\cal A}+\frac{7}{3}\;{\cal A}^2+\frac{\hat{f}_1}{f_1}\;{\cal A}\\
   \dot S&=&\frac{3}{4}\;S^2+p-\frac{7}{9}\;{\cal A}^2-\frac{4}{3}\;f_0\;\sqrt{\cal A}+\left(f_1-\frac{f_0^2}{4}\right)\;\frac{1}{\cal A}\\
   \hat S&=&-\left(\theta+\frac{3}{2}\;S\right)\;\left(\frac{2}{3}\;{\cal A}+\frac{f_0}{\sqrt{\cal A}}\right)
   \eea 
  From eqs (\ref{eqA}) and  (\ref{eqk}) we see that  ${\cal A}$ and ${\cal K}$ evolve in time in   opposite way, so that,   if ${\cal A} $ is growing, ${\cal K}$ is decreasing or vice versa.
  Analogous behaviour happens for the thermodynamical quantities, density and 
  temperature such that: $\dot n=-\theta\;n $ and $\dot T=\theta\;T $.
  \bea
  &&\dot{\hat{T}}=(\hat{\theta}+{\cal A}\;\theta)\;T+\left(\frac{2\;\theta}{3}-\Sigma\right)\;\hat{T},\qquad
  \dot{\hat{n}}=-(\hat{\theta}+{\cal A}\;\theta)\;n-\left(\frac{4\;\theta}{3}+\Sigma\right)\;\hat{n}\\&&
  \dot{  (T\;n)}=0,\qquad \dot{ \widehat{(T\;n)}}=-\left(\frac{ \theta}{3}+\Sigma\right)\;\widehat{(T\;n)}
    \eea
%  \\
%  There are special values  for some  form factors that deserve special considerations.
% To evade such intrinsic instability we can fix the following dynamical points (always for ${\cal A}\neq 0$)
% \\
% %\subsection{Case $S=0$ and static LRS space time }
%   For $S=0$ (i.e. $\Sigma=\frac{2}{3}\;\theta$) we have two options \footnote{ From $\hat S+\frac{3}{2}\;\phi\;S+\theta\;\phi=0\to \theta\;\phi=0$}, one with $\phi=0$ and another with $\theta=0$:
% 
% \begin{itemize}
% \item When $\phi=0$ the eqs reduce to
%     \bea
%     {\cal K}=\rho=- p,\quad \dot\rho=\hat\rho=  \dot{\cal A}=0,
%    \quad      \dot \theta=\hat{\cal A}+{\cal A}^2+\rho-\theta^2,\quad \dot{\hat{ \!\!\cal A}} =-\theta\;\hat{\cal A}
%    \eea
%    from which, deriving   the eq for $\theta$, we get a self contained evolution  
%    \be
%    \ddot{\theta}+3\;\theta\;\dot\theta+\theta\;(\theta^2-{\cal A}^2-\rho)=0
%    \ee
%    with a solution $\theta^2={\cal A}^2+\rho$.  
%    \footnote{We check the solution ${\cal A}=n=0$, $R= $ const , $F=1$,  
%    $\theta^2=\rho=\Lambda=-\frac{\k}{y_0^2}$
%        in   coordinate formalism (\ref{ca}). }
%    
%    \item When $ \theta=0$ we have a static and spherically symmetric space time where $\dot\rho=\dot n= \dot\phi=  \dot{\cal A}=0$ and 
% \bea
%% \hat p=n\;{\cal A},\quad \hat{\cal K}=-\phi\;{\cal K},\quad
%   {\cal K}= - p+\phi\;{\cal A}+\frac{\phi^2}{4},\quad  \hat \phi= \phi\;{\cal A}-\frac{\phi^2}{2}-p-\rho ,
%   \quad \hat{\cal A}=-{\cal A}^2-\phi\;{\cal A}+\frac{3\;p+\rho}{2}
%   \eea
%   Note that in this case also    $\Sigma=0$. 
%   \end{itemize}    

% \newpage
 \subsection{Null Geodesics}\label{ng}
   Due to the fact that cosmological observations rely on the detection of  the light emitted from far away sources, a non homogeneous background makes thinks more subtle for the mixing in between temporal and spatial variations.
   Moreover while the Einstein eqs are traced by the matter geodesic vector $u^\a$, 
   the geometry of light propagation  is dictated by the tangent vector $k^\a=d x^\a/d\nu$ (with $\nu$ the affine parameter). In the framework of
   geometrical optics,
   %is the base for the study of light propagation with  wavelength  much shorter than both the local curvature
  % radius and the typical scale over which the amplitude and the wavelength change appreciably.
    null geodesics  eqs are given by 
\bea
k^\a\;k_\a=0,&& \quad (k^\a)'\equiv\frac{\delta k^\a}{\delta\nu}=
k^\b\;\nabla_\b\;k^\a=0
%,\qquad
%\\&&
%E\;(\dot k^\a+\k\;\hat k^\a+\k^\b\nabla_\b\;k^\a)=0
\eea
The photon momentum %is given  by a null (geodesic) vector $k^\a$ it 
can be decomposed along the $u^\a$ vector as
  \be
  k^\a=E\;(u^\a+e^\a)
  \ee
   where $e^\a$ is a space like vector ($e^\a\, e_\a=1,\;e^\a \,u_\a=0$) that defines the direction of the photon  and $E=-k^\a\,u_\a$ is 
   the energy of the photon  relative to the observer defined by the vector $ u^\a$. 
The energy evolution along the null geodesic is given by the derivative of $E$ along $\nu$
\bea\label{eqE}
  &&\frac{\partial E}{\partial \nu}=k^\a\,\nabla_\a \,E=-k^\a\,k^\b\;\nabla_\a\,u_\b=
 -E^2\;\left(\frac{\theta}{3} +\sigma_{\a\b}\;e^\a\;e^\b+{\cal A}_\a\;e^\a  
\right)
\eea
where we applied the decomposition (\ref{gradu}).
The redshift $z$ of a source is defined as the observed photon wavelength divided by the wavelength at the source  minus one, and being the wave length inversely proportional to $E$ we get
 \be
    1+z\equiv \frac{E_G}{E_O}= \frac{(u_\a\;k^\a)\,|_G}{(u_\a\;k^\a)\,|_O}
    \ee
   where $u_\a$ is the four-velocity of the perfect fluid evaluated  for the galaxy (G) and for the observer (O).
Then we normalised the  null affine parameter with $\nu=0$ at the observational point   setting $dz=dE$.
With (\ref{eqE}) we get the evolution of the redshift along the light cone (specialised to a LRS space time) :
    \bea\label{dzn}
&& \frac{dz}{d\nu}=-(1+z)^2\;\left(\frac{\theta}{3} +{\cal A}\;v_\a\;e^\a+\Sigma\;(v_{\a}\;e^\a)^2\right)
%\;\;\;{\rm on\; a\; LRS \;space\;time}
%E\;\left(\dot E+\k\;\hat E +\k^\a\,\nabla_\a\,E\right)
\eea
Where we see the various cosmological contributions  to the redshift of the photon coming from the expansion rate , the acceleration of the observer and the shear.
 \\It is clear that redshift measurements are observations on null cones that sample the radial direction. 
 Also  the two dimensional orthogonal space directions (the screen space)   experience the expansion of the space time, but in a non homogeneous space it is important to distinguish such effects.
 So we introduce the {\it observed \underline{radial Hubble} rate} (line of sight expansion) as
 \footnote{${\cal H}_{\parallel}$ can be determined by measurements of standard ruler (as BAO) or standard chronometers (as the differential ages of ancient elliptical galaxies) \cite{Jimenez}}
 \be\label{Hpar}
 {\cal H}_{\parallel}=\frac{1}{3}\;\theta+\sigma_{\a\b}\;e^\a\;e^\b-{\cal A}_\a\;e^\a=-\frac{1}{(1+z)^2}\;\frac{dz}{d\nu}
 \ee
   versus the {\it  \underline{orthogonal Hubble} rate} (transverse expansion rate)
   \be\label{Hper}
   {\cal H}_{\perp}=\frac{1}{3}\;\theta-\frac{1}{2}\;\sigma_{\a\b}\;e^\a\;e^\b%-{\cal A}_\a\;e^\a
   \ee
   Finally {\it the \underline{volume expansion rate}} that gives the   usual definition of the Hubble parameter, i.e. the expansion of the  volume parameter $\theta$, is given by
   \be\label{Hvol}
   \theta={\cal H}_{\parallel}+2\;{\cal H}_{\perp}+{\cal A}_\a\;e^\a,\qquad {\cal H}=\frac{\theta}{3}=\frac{\dot l}{l}
\ee 
where $l$ is a representative length given by 
$dV=\eta_{\m\a\b\gamma}dx^{\m}\,dx^{\a}\,dx^{\b}\,dx^{\gamma}=\sqrt{det|^{(3)}g_{\a\b}|}\;d^3x$   and $l^3=\sqrt{det|^{(3)}g_{\a\b}|}$.
The presence of different expansion rates in different directions  is at the basis of 
  of the Alcock-Paczynski \cite{Alcock:1979mp} test in a general spacetime. 
  For an object that is known to be spherically symmetric, the ratio between his observed angular size over the radial extent in redshift space is a function of the redshift and the space time geometry.
   An isotropic Hubble rate clearly implies ${\cal H}_{\perp}={\cal H}_{\parallel}$ that is the case in a FRW model. 

\section{Coordinate approach, Lema$\hat{\rm i}$tre metric}
\label{ca}

To obtain   explicit solutions we need  to connect the above formalism directly to the metric components.
As described in detail in \cite{vanElst:1995eg} (see also \cite{bolejko} for applications to shell crossing), for a LRS space
we can use the following local metric in $(t, r, x, y)$ coordinates:
\be\label{g}
ds^2=-F^2(t,r)\;dt^2+X^2(t,r)\;dr^2+R^2(t,r)\;(\;dx^2+D(x,\;\kappa)\;dy^2)
\ee
where $D(x,\;\kappa)=(\sin x,\; x,\;\sinh x)$ for $\kappa=(1 \;{\rm spherical},\;0 \;{\rm euclidean},\;-1\;{\rm hyperbolic })$, that describes the geometry of the 2-$d$ sheet and $R(t,r)$ is the {\it area radius} coordinate. 
For a spherically symmetric inhomogeneous fluid  ($\kappa=1$) we have the so called 
{\it Lema$\hat{\rm i}$tre} metric \cite{Lemaitre} %that describes a spherically symmetric inhomogeneous fluid
 while in the  special case of dust ($p=0$) with a cosmological constant, the above metric reproduces the {\it Lema$\hat{\rm i}$tre-Tolman} (LT) model
 \footnote{It is important to stress the difference with the huge literature present for the LT models. 
 At background level, as already show in Tab \ref{tab1}, LT models have null acceleration ${\cal A}=0\to \partial_r p=0\to F^{(0,1)}(t,r)=0\to \Lambda'(r)=0$, so their metric can be written as
 \be\label{ltm}
ds^2=- dt^2+\frac{(R^{(0,1)}(t,r))^2}{1+ E(r)}\;dr^2+R^2(t,r)\;(\;dx^2+D(x,\;\kappa)\;dy^2)
\ee
with $E(r)$ an arbitrary function of integration proportional to  the curvature of space at each $r$ value  (to be compared with (\ref{ltX})). The time independence of $E$ allows an explicit solution for eq (\ref{eqY}).}.
Within the Lema$\hat{\rm i}$tre models the coordinates %$x^\m$
 are comoving with matter flow and for a perfect fluid both 
 pressure and energy density are functions of $t,\;r$ variables. 
The matter flow has four velocity  $u^\a=(1/F,\;0,\;0,\;0)$ while the special space four vector is $v^\a=(0,\;1/X,\;0,\;0)$.
The {\it centre} of the space is given by the solutions of $R(t,r)=0$.
% the isotropy group is 1-dim and the isometry group is $G_3$ (3-dim) acting on a space like 2-dim surface orthogonal to the vectors $u^\a$ and $v^\a$.
The covariant derivatives for a scalar function $f(t,r)$ become (\ref{dirD})
\be
\dot f\equiv \frac{f^{(1,0)}}{F},\qquad \hat{f} \equiv \frac{f^{(0,1)}}{X}
\ee
%Both pressure and energy density are functions of $t,\;r$ variables and the perfect fluid four velocity is $u^\a=(1/F,\;0,\;0,\;0)$ while the special space four vector is $v^\a=(0,\;1/X,\;0,\;0)$. 
Note also that in many articles the $X(t,r)$ function is written as
\be\label{ltX}
X(t,r)\equiv \frac{R^{(0,1)}(t,r)}{\sqrt{1+E(t,r)}}\;\to\; E =\hat R^2-1
\ee
where the  $E$ function ($>-1$ ) corresponds to the curvature parameter \cite{Lasky:2010vn}.

The geometrical quantities in ${\cal D}_2$,  (\ref{subD}), are given by
\footnote{In a conformal metric of a flat FRW we have that 
%the only non zero form factor are
$\theta=-\frac{2}{3}\;S=3\frac{\dot a}{a}=3\;\frac{ a'}{a^2},\;\;\phi=\frac{2}{r\;a}$.}
\bea
&& {\cal A}=\frac{\hat F}{F}%=\frac{F^{(0,1)}}{F X}
,
\qquad \phi = \frac{2\;
  \hat R }{  R}
%  =\frac{2\; Y^{(0,1)}}{X\; Y}
  ,\qquad
   \mathcal{K}=
   \frac{ \kappa}{R^2} ,\qquad \xi = 0,\qquad \Omega = 0
   \\ &&\nonumber
   \;\;
 \Sigma=\frac{2}{3}\left(\frac{\dot X}{X}-\frac{\dot R}{R}\right)
 %=\frac{2  } {3 \;F  }\left(  \frac{  X^{(1,0)} }{  X}-\frac{  Y^{(1,0)} }{  Y}\right)
 ,\qquad
   \theta = \left(\frac{\dot X}{X}+2\;\frac{\dot R}{R}\right),\qquad S=-2\;\frac{\dot R}{R}
   \\ &&\nonumber
   {\cal E}=\frac{1}{3}\left(
  - {\cal K}+\frac{\ddot R}{R}- \frac{\ddot X}{X}+\frac{\hat{\hat F}}{F}-\frac{\hat{\hat R}}{R}+
  \left(\frac{ {\hat R} }{R}\right)^2-
  \left(\frac{ {\dot R}  }{R }\right)^2+\frac{{\dot R}\;{\dot X}}{R\;X}-\frac{{\hat R}\;{\hat F}}{R\;F}
   \right)
   % =\frac{1}{F}\left(\frac{X^{(1,0)}}{X}+\frac{2\;Y^{(1,0)}}{Y}\right),
\eea
and the two Hubble rates (\ref{Hpar}, \ref{Hper}) result
\bea\label{H12}
{\cal H}_{\parallel}=\frac{\dot X}{X}-\frac{\hat F}{F}\qquad {\rm and} \qquad {\cal H}_{\perp}=\frac{\dot R}{R}
\eea
Note that the scalar shear $\Sigma$ is proportional to the difference between the radial and azimuthal expansion rates.
%The vorticity is zero so it   implies both the existence of a global cosmic time and a 3-dim global spacelike hypersurfaces orthogonal to the fluid congruence.

\subsection{The N$\Lambda$CDM model}
Applying the above eqs to the N$\Lambda$CDM model 
 we can integrate  the density and the EMT conservation  equations  
 \bea\label{eqL}
&&\dot \Lambda=0\;\;\to\;\; \Lambda=\Lambda(r)\\\label{eqn}
&&\frac{\dot n }{n}+ \;\left(\frac{ \dot X }{X}+\frac{2  \;\dot
   R }{R}\right)=0\;\;\to\;\; n=\frac{\mathbf{n}(r) }{X\; R^2}\\ \label{eqx}
&&    n  \; \frac{\hat F }{F}= \hat \Lambda \;\;\to\;\; 
   X= \frac{ \;\mathbf{n}  }{R^2\;
   \Lambda'}\;\frac{F^{(0,1)}}{F},\qquad \Lambda'\neq0
\eea
where $\Lambda\equiv\Lambda(r)$ and $\mathbf{n}\equiv\mathbf{n}(r)$ are effectively introduced as spatial  boundary conditions.
%Note that in eq.(\ref{eqx}) we assumed $\Lambda'\neq 0$. % so, from now on, we cannot set it to zero.
Then we introduce the Misner-Sharp mass (see also \cite{Marra:2011zp}, \cite{Lemaitre}, \cite{Hellaby})
\be\label{eqdM}
M(t,r)\equiv
%\frac{ Y}{2}\;   \left(1+\frac{\left(Y^{(1,0)}\right)^2}{F^2}-\frac{\left(Y^{(0,1)}\right)^2}{X^   2} \right)=
   \frac{ R}{2}\;
   \left(1+  \dot R^2  -  \hat R^2  \right)=\frac{ R}{2}\;
   \left(   \dot R^2  -  E  \right)
\ee that, in the Newtonian limit, represents  the mass inside the shell of radial coordinate $r$.
%Our intent  is to get  a close equation for the $R(t,r)$ function. 
Using the $M$ variable  the Einstein eqs read
\footnote{\it From eq.(\ref{eqpr}) the infinite density limit is obtained when $\hat R= 0$ with $R\neq 0$ (\ref{eqYh}).
 These singularities correspond to shell-crossing singularities or caustics  (i.e. cusps for the dark matter distribution).
%  This is akin to fluid shock waves.
  It is a generic prediction that the evolution for an absolutely cold dark matter system from near uniform initial conditions under the effect of gravity, at non linear level, bring to generation of caustics. %  (as it happens also for similarity solutions in cold spherical infall).
In general  it is expected that,  on the smallest scales,  
the fluid matter model, that results a macroscopic
approximation for the smooth behaviour of matter fields, is not appropriate for
the study of highly non linear gravitational phenomenon.
%In general such a singularities are not considered to be so serious.
 %In fact it is reasonable   that the fluid matter model, that results a macroscopic
%approximation for the smooth behaviour of matter fields, is not appropriate for
%the study of gravitational phenomenon on the smallest scales.
For realistic dark matter candidates it is expected that such a 
 divergences can be tamed for example by  thermal effects and it
 results as a  motivation for  alternatives to CDM (warm dark matter, fuzzy dark matter, dust replaced by the Vlasov model, inviscid fluid replaced by viscous).
  In Tolman models (where $p=0$) the necessary and sufficient conditions which ensure no shell crossings  are described in \cite{hellab}.
 }
\be\label{eqpr}
\rho=\Lambda+n=%2\;\frac{M^{(0,1)}}{Y^2\;Y^{(0,1)}}=
2\;\frac{\hat M }{R^2\;\hat R },\qquad 
p=-\Lambda=%-2\;\frac{M^{(1,0)}}{Y^2\;Y^{(1,0)}}=
-2\;\frac{\dot M }{R^2\;\dot R }
\ee
The   pressure equation can be  integrated giving
\be\label{eqM}
 M= \frac{1}{6}\;\left(\Lambda  \;R^3 \;+\mathbf{m}(r) \right)
\ee
where   $\mathbf{m}\equiv\mathbf{m}(r)$ is another   spatial  boundary conditions.
Inserting  (\ref{eqM}) in (\ref{eqdM})  we get 
\bea\label{eqY}
%\dot R^2&=&-1+\hat{R}^2+\frac{\Lambda}{3}\;R^2+ \frac{\bf m}{3\;R}
%\\
{\cal H}_{\perp}^2\equiv \left(\frac{\dot R}{R}\right)^2&=&\boxed{\left(\frac{R^{(1,0)}}{F\;R}\right)^2=\frac{1}{3}\;
\left( \!\!\underbrace{\Lambda }_{DE} \!\!+ \underbrace{ \frac{\bf m}{ R^3}}_{DM} \!\! +  \underbrace{\frac{3\;E}{R^2}}_{Curvature} \right)}
\eea
while the radial derivative of $R$, using the expression for the energy density (\ref{eqpr}) and the density (\ref{eqx}), can be written  as
 \bea\label{eqYh}
 \hat{R}=\frac{X}{3\;{\bf n}}\;\left( \hat{\bf m}+\hat{\Lambda}\;R^3\right)=\frac{1}{3\;{\bf n}}\;\left(    {\bf m}'+{\Lambda}'\;R^3\right)=\sqrt{E+1}
 \eea
The same expression can be used also to  relate $n$ and $\bf{m}$ as
\be\label{mn}
\hat{\bf{m}}=n\;R^2\;\left(3\;\hat{R}-\frac{\hat F}{F}\;R \right)
\ee
so that, in general,  $n\neq0$  (or $\bf n\neq 0$) implies $\bf{m}\neq0$.  
\\
To have a flat FRW space time we need the following conditions: 
\begin{itemize}
\item $R=r \;a(t)$,  $F=a(t)$,  $X=a(t)$ so that ${\cal H}_{\perp}=\frac{\dot a}{a}$ and $\hat{R}=1$
 \item Boundary conditions for the matter content implies ${\bf m}=  3\,{  H}_0^2\;\Omega_m\;r^3$ and (\ref{mn}) requires 
 ${\bf n}=\frac{{\bf m}'}{3}=3\,{ H}_0^2\;\Omega_m\;r^2$   
  \item  $\Lambda =3\, {  H}_0^2 \;\Omega_{\Lambda}$ constant  in space.
  \end{itemize}
While the matter conditions can be easily implemented, the constancy of $\Lambda$ is contrary to our assumptions. 
 Following \cite{Lasky:2010vn}, we see that the general structure of the Local Hubble parameter  
 $ {\cal H}_{\perp}(t,\,r)$
    is quite similar to the FRW Hubble equation and
 allows to identify various contributions \footnote{ We note as 
 ${\cal H}_{\parallel}+{\cal A}={\cal H}_{\perp}\left(\frac{R \;R^{(1,1)}}{R^{(0,1)} \;R^{(1,0)}}-\frac{3 R^3 \Lambda
   '}{\mathbf{m}'+R^3 \Lambda '}\right)
  $ doesn't show such 
 a similarity  with the FRW Hubble equation.}
\bea
&& \Omega_\Lambda(r)=\frac{\Lambda}{3\;H_0^2},\\\nonumber
&& \Omega_M(t,\,r)=\frac{\bf m}{3\;H_0^2\;R^3},\\\nonumber
 &&\Omega_c(t,\,r)= \frac{E}{H_0^2\;R^2}=\frac{\hat{R}^2-1}{ H_0^2\;R^2}=
 \frac{1}{H_0^2\;R^2}\,\left(\frac{(R^3\;\Lambda'+{\bf m}')^2}{9\; {\bf n}^2}-1\right)
=\\
&& \qquad
 \frac{1}{3\;H_0^2}\;\left(
 \left( \left( \frac{{\bf m}'}{\bf n}\right)^2-9\right)\;\frac{1}{R^2}+  \frac{ \Lambda' }{{\bf n}^2}\;R\;\left(
  2 \;{\bf m}' + \Lambda' \;R^3\right)\right)
\eea
with the constraint $\Omega_\Lambda(r)+\Omega_M(t_0,\,r)+\Omega_c(t_0,\,r)=1$ and
where the current Hubble rate is $ H_0(r)=
{\cal H}_{\perp}(t_0,\,r)$.
The structure of $\Omega_c$ contains a term, proportional to $ \left(\left( \frac{{\bf m}'}{\bf n}\right)^2-9\right) $, that depends only on   dark matter structure while the second term, proportional to $\Lambda'$,   disappears for a true constant CC where $\Lambda' \to 0$.
\\
In order to have a close differential eq for only one component of the metric    we can rewrite  eq.(\ref{mn}) as
\bea
\label{eqF}
\frac{\hat{F} }{F}&=& \frac{3\;\hat{\Lambda} \; R^2 \;\hat{R}  }{ \hat{\mathbf{m}} +  \hat{\Lambda}
  \;R^3}
\qquad\to\qquad
\boxed{
\frac{F^{(0,1)}}{F}= \frac{3\;\Lambda '\; R^2 \;R^{(0,1)} }{ \mathbf{m}'+  \Lambda '\;R^3
 }}
 \eea
 so that $X$ (from (\ref{eqx})) is completely determined from the $R$ dynamics.
 Eq.(\ref{eqF}) and eq.(\ref{eqY}) represent a close system of partial differential equations governing the dynamics of the $F $ and $R$ functions. In Appendix we give the  self contained  evolution equation for the  $R$ function (\ref{eqy11}).
   The rest of kinematical quantities are here specified (for $\theta$ and $\Sigma$ we don't use eq (\ref{eqy11}) to have a simplest  expression)
   \bea
  && {\cal K}=\frac{1}{R^2},\qquad
    {\cal A}=\frac{ R^2\;
   \Lambda'}{ \mathbf{n} }\\\nonumber
 &&  \phi=\frac{2 \left( \mathbf{m}'+R^3\; \Lambda
   '\right)}{3\; \mathbf{n} \;R},\qquad  
  {\cal E}=\frac {  \mathbf{m}'+R^3 \;\Lambda '}{9\;R^2\;R^{(0,1)}}-\frac{\bf m}{3\;R^3}\\\nonumber
  && \theta\;F= \frac{R^{(1,1)}}{R^{(0,1)}}+\frac{2\;
   R^{(1,0)}}{R}-\frac{3 \;\Lambda
   '\;R^2 \;R^{(1,0)} }{  \mathbf{m}'+R^3\; \Lambda '}
   \\\nonumber
 &&  \Sigma\;F=  \frac{2}{3}\;
   \left(\frac{R^{(1,1)}}{R^{(0,1)}}-\frac{R^{(1
   ,0)}}{R}\right)-\frac{2 \;\Lambda
   '\;R^2\; R^{(1,0)} }{  \mathbf{m}'+R^3 \;\Lambda '}
   \eea
  From the non perturbative eqs (\ref{kf}) we can relate  the various space dependent integration constant 
  \bea
  {\cal K}=\frac{f_1}{\cal A}\;\to\; f_1=\frac{\Lambda'}{\bf n},\quad 
  \phi=\frac{2\;{\cal A}}{3}+\frac{f_0}{\sqrt{|{\cal A}|}}\;\to\;
   f_0=\frac{2\;{\bf m}'\;\sqrt{|\Lambda'|}}{3\;{\bf n}^{3/2}} 
  \eea
   Note that for $ \mathbf{n} =\;\frac{\mathbf{m}'}{3}$ we have 
   $f_0=2\;f_1^{1/2}$.

 \section{Analytical Approximations}
 We assume a spatial background distribution of  DM compatible with the FRW conditions  $ \mathbf{n} =\;\frac{\mathbf{m}'}{3}$. Then  eqs (\ref{eqF}) and (\ref{eqY}) give the closed system
  \bea \label{eqNM}
\left\{\begin{array}{lc}
\frac{F^{(0,1)} }{F }= \frac{3\;\Lambda '(r)\; R ^2 \;R^{(0,1)}  }{ \mathbf{m}'(r)+  \Lambda '(r)\;R^3 }
 ,\\ 
 \\
   X = \frac{\mathbf{m}'(r)\; R^{(0,1)}  }{\Lambda'(r)\;
   R ^3+\mathbf{m}'(r) } 
   \\
   \\
   \frac{R^{(1,0)} }{F \;R } = \sqrt{\frac{\Lambda(r)}{3
   }+\frac{\mathbf{m}(r)}{3\;   R ^3}+
   \frac{\Lambda'(r)\;R }{ \mathbf{m}'(r)^2}\;
   \left( R ^3 \;\Lambda'(r) + 2\;  \mathbf{m}'(r)\right)
   }
\end{array}
\right.
 \eea
that we  approximatively  solve  in different contexts.
In chapter (\ref{chFRW}) we  perturbatively solve the above eqs  expanding  around an homogeneous FRW space time  with a small, space dependent, correction to the CC.
% For specific values of the CC   eq (\ref{eqF}) ca be  exactly solved as analysed in chapter \ref{chF}.
% \\
Instead in chapter (\ref{chY}) we  solve  the above system   expanded for small $r$ values.
 %In all the above analysis we will give the structure of the metric (\ref{g}) while 
 Finally in  chapter  (\ref{geo}) we  study the light ray propagation with the Lema$\hat{\rm i}$tre metric using the above solutions.% for the different approximations.

  \subsection{Solutions  around an approximated FRW model }\label{chFRW}
  
  The exact   FRW solution for the  eqs (\ref{eqNM}) are obtained for
  \be
 {\rm FRW \;limit:} \quad \Lambda'=0,\quad
 {\bf m}(r)=3\;{  H}_0^2\;\Omega_m\;r^3,\quad{\bf n}(r)=\frac{{\bf m}'(r)}{3}=3\;{  H}_0^2\;\Omega_m\;r^2 
  \ee giving $F=1$, $R= a(t)\,r$.
  The scale factor $ a(t)$, in a De Sitter-Matter dominated universe, is
  % obtained from the following evolution equation
\be\label{sfrw}
\bar{H}(t)^2\equiv\left(\frac{ a'(t)}{ a(t)}\right)^2={  H}_0^2\;\left( \Omega_\Lambda+\frac{\Omega_m}{ a(t)^3}\right)\;\to\;  a(t)=\left(\frac{\Omega_m}{\Omega_\Lambda}\;
\sinh^2\left(\frac{3}{2}\;t\;{\cal H}_0\;\sqrt{\Omega_\Lambda}\right)\right)^{1/3}
\ee
We perturb such a configuration with a small space-dependent correction $\lambda(r)$ to the CC
\bea\label{eqll}
&&\Lambda(r)=3\;{ H}_0^2\;\Omega_\Lambda\;\left(1+\lambda(r)\right), \qquad \lambda(r)\ll1
%\\
%&& {\bf m}(r)=3\;{\cal H}_0^2\;\Omega_m\;r^3\\
%&& {\bf n}(r)=\frac{{\bf m}'(r)}{3}=3\;{\cal H}_0^2\;\Omega_m\;r^2
\eea
and we expand  the metric coefficient  at order ${\cal O}(\lambda^n)$ as
\bea\label{exp}
 %&&  Z(t,r)=\a(t)\;\sum_{n=0}\;\frac{Z_n(t,r)}{n\, !} ,\quad Z=F,\;R,\;X
  A=1+A_1+\frac{A_2}{2}+...,\quad A\equiv F,\,X ;
  \qquad 
   R= a(t)\;r\;\left(1+R_1+\frac{R_2}{2}+...\right).
     \eea
  For $F(t,r)$ we get, at second order,
 \footnote{  Where we use $\,
   _2F_1\left(-\frac{2}{3},1;\frac{5}
   {6};x\right)\equiv h_1(x),\;\,
   _2F_1\left(\frac{1}{3},1;\frac{11}
   {6};x\right)\equiv h_2(x)$ and $x=-a(t)^3\;\frac{\Omega_{\Lambda}}{\Omega_m}$.}
   % checking the convergence region at the end,   finding that
     \bea\label{eqfrwF}
    F &=&1+
    \frac{\Omega_\Lambda}{\Omega_m}\, a(t)^3\,\lambda(r)
     +\frac{  \Omega _{\Lambda }^2 }{\Omega _m^2} \, a(t)^6\,\lambda
   (r)^2
     -
    % \\ &&\nonumber
     \frac{ \Omega _{\Lambda } \left(  h_2(x)  -1\right)   }{6 \,{  H}_0^2\; \Omega _m^2}
    a(t)^4 \left(\lambda '(r)^2+4\; \mathbb{L}_2(r)\right)  
        +...
     \eea
   where $\mathbb{L}_2(r)=\int^r_0dx\;\frac{\lambda'(x)^2}{x}$. 
   For the components of $R(t,r)$ and $X(t,r)$ we get
      \bea \label{eqfrwR}
      R &=&a(t)\;r\;\left(1+\frac{ \Omega _{\Lambda } }{3\; \Omega _m}
       a(t)^3\;\lambda (r)-\frac{ 
  h_2(x) %_2F_1\left(\frac{1}{3},1;\frac{11}{6};-\frac{a^3 \Omega _{\Lambda }}{\Omega_m}\right)
   -1 }{3 \;{ H}_0^2 \; \Omega _m}\; a(t)\; \frac{\lambda '(r)}{r}+\frac{R_2(t,r)}{2}+...
   \right)
      \eea
     \bea\label{eqfrwX}
     X &=& a(t) \left(1+\frac{ \Omega _{\Lambda } }{3 \,\Omega _m}\, a(t)^3\,\lambda (r)+
    % \right. \\ &&\nonumber \left. 
     \left(\frac{5 \left(1- \,   h_1(x)    \right)}{ a(t)^2\, \Omega _{\Lambda }}
    +\frac{4 \, a(t)}{\Omega _m}\right)\frac{\lambda ''(r)}{12 \,{  H}_0^2}+...\right)
     \eea
     while for the other geometrical quantities we have
     \bea
    && {\cal A}(t,r)=\frac{ \Omega _{\Lambda }\;  a(t)^2 \; \lambda '(r)}{\Omega _m}
     \\&&\label{Hr}
     {\cal H}_{||}(t,r)= \bar{ {H}}(t) \left(1+ \frac{\lambda ''(r)}{H_0^2}\; \left(
     \frac{a(t)}{\Omega _m}\, \left(\frac{5
     \; h_1 '(x)}{4}+\frac{1}{3}\right)+\frac{5 \;
   ( h_1(x)-1)}{6 \;a(t)^2 \; 
   \Omega_{\Lambda }}\right)\right)-{\cal A}(t,r)
     \\&& 
     {\cal H}_\perp(t,r)= \bar{  {H}}(t) \left( 1+ 
     \frac{\lambda '(r)}{H_0^2\, r}\,\left(\frac{a(t)^4\; \Omega _{\Lambda }\;
   h_2'(x)}{\Omega _m^2}+\frac{a(t)\;
   \left(1-h_2(x)\right)}{3\;
   \Omega _m}
     \right)\right) 
          \eea
     The convergence region of our series expansion can be inferred looking  to eq (\ref{eqfrwF})
     \be
     \frac{\Omega_m}{\Omega_\Lambda}\; a(t)^3\;\lambda(r)\ll1\qquad {\rm and\;today}\;\;\; \lambda(r)\ll  \frac{\Omega_\Lambda}{\Omega_m}\simeq 2
     \ee
     %Imposing $\a(t=0)=0$ we have $\a(t_0)=1$ for $\tau=Arcsinh(\sqrt{2})\simeq1.146$. 
     Then  analysing  the solution for $R$  (\ref{eqfrwR})  and $X$ (\ref{eqfrwX}) we need also the present perturbativity constraints
     \bea\label{l1r}
     \lambda(r),\;\frac{\lambda'(r)}{ r\;{ H}_0^2},\;\frac{\lambda''(r)}{ {  H}_0^2}\ll1
     \eea
     or, in  the future ($a>1$), we can perturbatively reach the region with
     \be
      a(t)\ll {\rm Max} \left[\frac{1}{\lambda^{1/3}},\;\frac{r\;{  H}_0^2}{\lambda'},\; \frac{ {  H}_0^2}{\lambda''}\right]
     \ee
     Note that (\ref{l1r}) must be valid for all $r$ values  so, for example,
       in the limit $r\to 0$ we need an expansion of the form 
       $\lambda(r)\sim  \lambda_2\;r^2 +...$ without the linear term and the constant term that can be absorbed  inside $\Omega_{\Lambda}$
       (the structure of the above expansion is confirmed also in the next chapter).
       At present time $a(t_0)=1$ and at the center of the vacuum bubble $r=0$ from (\ref{Hr}) we get
       \bea\label{zeror}
       {\cal A}(t_0,r=0)=0,\quad {\cal H}_{||}(t_0,r=0)=H_0\;
       \left(1+0.667\;\lambda''(0)\right),\quad {\cal H}_\perp(t_0,r=0)=H_0
       \eea
      Conversely, far away $r\to \infty$, we impose that  $\lambda(r)$ is approaching a constant $ \lambda_0$ (that can be zero) with $\lambda'(r),\;\lambda''(r)\sim 0$ (the space gradients die earlier) so that
        \bea\label{infr}
       {\cal A}(t,r\gg1/H_0)\sim  0,\quad {\cal H}_{||,\perp}(t, r \gg 1/H_0)\sim  \bar H(t) 
              \eea
              matching the FRW background solution (\ref{sfrw}).
 We stress that the above eqs (\ref{eqfrwF},\ref{eqfrwR},\ref{eqfrwX}) are perturbative solutions of the background equations (\ref{eqNM}) and not perturbations  of the Lema$\hat{i}$tre metric (\ref{g}).
 In this approximation, from the eqs (\ref{zeror},\ref{infr}), we identify $H_0$ with the Hubble Planck data, see (\ref{Hpl}).
 
    \subsection{Solution in the small $r$ expansion limit}\label{chY}
   Explicit results from the   equations (\ref{eqNM}) can be obtained 
   from the solutions around our neighbour    universe. 
    Following ref.\cite{Partovi:1982cg},
   we perform a small $r$ expansion for all the form factors of the metric and associated quantities
   \bea\label{ser}
 &&  f(t,r)=\sum_{n=0}\;\frac{f_n(t)}{n\, !}\;r^n,\quad f=F,\;R,\;X,\;\eta,\;M\\
  && {\bf f}(r)=\sum_{n=0}\;\frac{{\bf f}_n}{n\, !}\;r^n, \quad {\bf f}=\Lambda,\;{\bf n},\;{\bf m}
   \eea
   Note that the coefficients of bold quantities  ${\bf f}_n$ are just numbers.
Matching order by order the continuity eqs and the Einstein equations (\ref{eqNM}), for spherically symmetric solutions ($\k=1$), we get the following relations:
   \begin{table}[htp]
\begin{center}
\begin{tabular}{||c||c|c|c|c|c||}
\hline\hline
&$F(t,r)$ & $X(t,r)$& $R(t,r)$& $\eta(t,r)$ & $\Lambda(r)$\\
\hline
$n$=0&1 & $a(t)$& $0$& $3\;\frac{{  \cal H}_0^2\;\Omega_m}{a(t)^3}$ & $\Lambda_0$\\
\hline
$n =1\to \times \;r$&0 & $ R_2(t)$& $a(t)$& $\frac{{\bf n}_3}{6\;a(t)^3}-6\;\frac{{ \cal   H}_0^2\;\Omega_m\;R_2(t)}{a(t)^4}$ & $0$\\
\hline
$n=2\to \times \;\frac{r^2}{2}$& $\frac{3\;{ \cal   H}_0^2\; a(t)^3\;\Omega_{\chi}}{2}\; $ & $-{ \cal   H}_0^2\;\Omega_{\chi}\;a(t)^4+R_3(t)$ & $ R_2(t)$ & $\eta_2(R_i)$
 & $\Lambda_2$\\
\hline\hline
\end{tabular}
\end{center}
\caption{The leading coefficients from the series (\ref{ser}) that solve the eqs of motion  for the different variables.}
\label{default}
\end{table}%
\footnote{$\eta_2(t)= -\frac{2\;
   \mathbf{n}_3 \;R_2(t)}{3\;
   a(t)^4}+\frac{\mathbf{n}_4}{12\;
   a(t)^3}+\frac{ { \cal H}_0^2\;\Omega
   _{\mathit{m}}}{2\;a(t)^5}\;
   \left( 33 \;R_2(t)^2  -10\;
   R_3(t) \;a(t)\right)+3\;
    { \cal H}_0^4 \;\Omega_{\mathit{m}} \;\Omega _{\chi}$}
   \\
At order ${\cal O}(r^0)$ the Hubble function is given by
   \bea
 &&  \bar{\cal H}(t)^2\equiv\left(\frac{a'(t)}{a(t)}\right)^2=\frac{\Lambda_0}{3}+\frac{{\bf n}_2}{6\;a(t)^3}+
 \frac{4\;\Lambda_2\;a(t)}{3\;{\bf n}_2}={  \cal H}_0^2\;\left(\Omega_\Lambda+\frac{\Omega_m}{a(t)^3}+\Omega_{\chi}\;a(t)
 \right) \label{Hrt}
 %\;\;\Lambda_1=0
\eea
after the identifications: $ \Lambda_0 =3\,{  \cal  H}_0^2\,\Omega_\Lambda,\;
 {\bf n}_2 =6\,{   \cal H}_0^2 \;\Omega_m,\;
  \Lambda_2  =9/2\;{ \cal   H}_0^4\;\Omega_m\;\Omega_\chi$.
  In such approximation for the background Hubble constant we use the symbol ${\cal H}_0$ that refers to the nearby Hubble constant measurement (\ref{Hnear}) contrary to the previous case (\ref{sfrw}) where it was used $H_0$ (Planck value).
Note the presence of an unusual $\Omega_{\chi}$ component whose eqs of state result $w_{\chi}=-\frac{4}{3}\simeq-1.33$.
The $ M(t,r)$ function (\ref{eqdM})  start his expansion at ${\cal O}(r^3)$ 
\bea
M_3(t)=3\;{ \cal  H}_0^2\,\left(\Omega_m+ \Omega_\Lambda\;a(t)^3\right)
%,\quad M_4=\frac{\bf m_4}{6}+18\;{  H}_0^2\;\Omega_\Lambda\;a(t)^2\;R_2(t)
\eea
From (\ref{eqpr}) we get
${\bf m}_i=3\;{\bf n}_{i-1}$ corresponding to ${\bf n}(r)=\frac{{\bf m}'(r)}{3}$. Finally  there are two functions $R_{2,3}(t)$ that satisfy a coupled system of evolution equations ($\Lambda_i\equiv 3\;{ \cal  H}_0^2\;{\ell}_i$) (\ref{R2},\ref{R3}).
  The leading terms for the kinematical variables result
   \bea \label{RAS}
 && ^{(3)}{\cal R}(t,r)=%\frac{6\;(\beta-3\;\gamma)}{a^2}=
   -6\;{ \cal H}_0^2\;\Omega_{\chi}\;a(t)%-\frac{4\;\Lambda_2\;a}{{\bf n}_2}
   ,\quad
   {\cal A}(t,r)= \frac{3\;{  \cal  H}_0^2}{2}\;\Omega_{\chi}\;a(t)^2\;r,
   %\frac{\Lambda_2\;a^2\;R}{{\bf n}_2},
   \quad
   \Sigma(t,r)= \frac{r}{3}\,\left(\frac{R_2(t)}{a(t)}\right)'
%   \left(\frac{Y_2'}{a}-\frac{a'\;Y_2}{a^2} \right)\;\frac{r}{3}
   ,\\\nonumber
  && \phi(t,r)=\frac{2}{a(t)\,r},\quad
  {\cal E}(t,r)=\left(\frac{{\bf n}_3}{6}- \frac{{\bf n}_2}{a}\,R_2(t)\right)\,\frac{r}{12\,a(t)^3},\quad \theta(t,r)=3\,{ H}(t)+6\;\Sigma(t,r) 
   \eea
   The above expansion result reliable as soon as
   \bea
   ( {\cal H}_0\,r)^2\,a(t)^3\,\Omega_{\chi},\;\frac{R_2(t)\,r}{a(t)}\ll1
   \eea 
   and at present time $a(t_0)=1$, $({ \cal H}_0\,r)^2 \,\Omega_{\chi},\; R_2(t_0)\,r \ll1$
     
  \section{Null Geodesics on a Lema$\hat{\rm i}$tre universe}\label{geo}
  
   Because the observation are done along the light cone, first of all  we have to study
 the null geodesic eqs  as a functions of the redshift.
  For a   source directed towards an observer located at 
 the symmetry centre of the model,  null geodesic are given by \cite{Partovi:1982cg}
\bea \label{eqzt}
(1+z)\;\frac{d t}{dz}=\frac{X}{F^{(0,1)}-X^{(1,0)}},\quad (1+z)\;\frac{d r}{dz}=-\frac{F}{F^{(0,1)}-X^{(1,0)}},\quad
\frac{d t}{dr}=-\frac{X}{F}
\eea
where $r=r(z)$ and $t=t(z)$. Concerning the photon geodesic eq.(\ref{dzn}), 
   being $\frac{dz}{d\nu}=E\;\frac{dz}{d\tau}= \frac{(1+z)}{F}\;\frac{dz}{dt}$ and $v_\a\;e^\a=\cos(\theta)$ 
   on a LRS space we get
   \bea\label{dztr}
%\frac{dz}{dv}=E\;\frac{dz}{d\tau}= \frac{(1+z)}{F}\;
\frac{dz}{dt}=
- (1+z)\;F\;\left(\frac{1+2\;\cos(\theta)^2  }{3}  \frac{\dot X}{X}+\frac{2\;(1-\cos(\theta)^2)}{3}\;\frac{\dot R}{R} +\frac{\hat F}{F}\;\cos(\theta) \right)
\eea
to be compared with the result of eq.(\ref{eqzt})
\bea\label{dzt}
\frac{dz}{dt}=(1+z) \;\left(\hat F-F\;\frac{\dot X}{X}\right) ,
\eea
So that exactly we have $\cos(\theta)=-1$. This corresponds to have the observer exactly at the centre of the space time ($r=0$) so that congruence are isotropic. For off centre observers, where $cos\theta\neq\pm 1$ it appears a dipole correction induced by the acceleration and a quadrupolar correction by the shear.
  There are many "distance" definitions  between two points in cosmology, measuring the separation between events on radial null trajectories we have:
   \bea \label{da}
&& {\rm Luminosity \;distance} \qquad d_L(t_0,\,z)=(1+z)^2\;R(t(t_0,\,z),\, r(t_0,\,z))
\\
&&\nonumber{\rm Comoving \;distance} \qquad d_C(t_0,\,z)=(1+z)\;R(t(t_0,\,z),\, r(t_0,\,z))
\\
&&\nonumber{\rm Angular\;diameter \;distance} \qquad d_A(t_0,\,z)= R(t(t_0,\,z),\, r(t_0,\,z))  
 \eea
 Following \cite{Bellido} we can work out some formulas inspired by the more familiar FRW models
 that can help understanding  the dynamics of the system. 
 For example we can build an effective equation of state along the radial ($\parallel$) or the 
 angular ($\perp$) directions as
 \bea\label{wra}
 w_{\parallel,\perp}=-1+\frac{2}{3}\;\frac{d \log {\cal H}_{\parallel,\perp}}{d\log(1+z)}
 \eea
 also  the effective acceleration can be a useful function
 \be\label{acc}
 q_{eff}(z)=-1+\frac{d \log {\cal H}_{\parallel}}{d\log(1+z)}
 \ee
  In the next chapters we will solve    eqs (\ref{eqzt}) in the almost FRW approximation (\ref{Gfrw}) and in the small $r$ limit (\ref{Gr}).
 
  \subsection{Geodesics  around an approximated FRW model }
  \label{Gfrw}
  The dependence on redshift for the Hubble parameter in a $\Lambda$CDM model in the last evolution period results
    \be\label{Hcdm}
   {\cal H}_{\Lambda CDM}(z) \equiv \bar{  H} (z)=H_0\;\sqrt{\Omega _{\Lambda }+(z+1)^3 \;\Omega _m}
 \ee
 Along the light geodesic instead of $t(z)$ we use $a(z)$ and implement the following 
 expansion at order ${\cal O}(\lambda^n)$ 
 \bea
  a(z)\equiv  \alpha_0(z)+\alpha_1(z)+..., \quad  a(z=0)=1\\
 r(z)\equiv r_0(z)+r_1(z)+..., \quad r(z=0)=0
 \eea
At zero oder  ${\cal O}(\lambda^0)$ we get the FRW solutions
 \bea
&&  \alpha_0(z)=\frac{1}{1+z}
 \\&& r_0(z)=\frac{2 }{5\, H_0\,
   \sqrt{\Omega _m}}
\;\left(\,
   _2F_1\left(\frac{1}{2},\frac{5}{6};\frac{11}{6}
   ;-\frac{\Omega _{\Lambda }}{\Omega
   _m}\right)-\frac{\,
   _2F_1\left(\frac{1}{2},\frac{5}{6};\frac{11}{6}
   ;-\frac{\Omega _{\Lambda }}{(z+1)^3 \Omega
   _m}\right)}{(z+1)^{5/2}}\right) \eea
   while at order ${\cal O}(\lambda)$  we have  two   coupled differential eqs to be numerically  solved
   \footnote{
   \bea
   f(z)=\frac{1}{6 \,H_0^2}\;\left(\frac{6 \; _2F_1\left(\frac{1}{3},2;\frac{11}{6};-\frac{\Omega _{\Lambda }}{(z+1)^3\,
   \Omega _m}\right)-2}{(z+1)^3 \Omega _m}-\frac{5 \left(\,
   _2F_1\left(-\frac{2}{3},1;\frac{5}{6};-\frac{\Omega _{\Lambda }}{(z+1)^3 \Omega
   _m}\right)-1\right)}{\Omega _{\Lambda }}\right)
   \\
   k(z)=\frac{ 1}{4 \,H_0^2\, \bar{  H} (z)} \left(\frac{5 \left(\,
   _2F_1\left(-\frac{2}{3},1;\frac{5}{6};-\frac{\Omega _{\Lambda }}{(z+1)^3 \Omega
   _m}\right)-1\right)}{\Omega _{\Lambda }}-\frac{4 \,
   _2F_1\left(\frac{1}{3},2;\frac{11}{6};-\frac{\Omega _{\Lambda }}{(z+1)^3 \Omega
   _m}\right)}{(z+1)^3 \,\Omega _m}\right)
   \eea}
   \bea
 &&  \alpha_1'(z)+
 \frac{\alpha_1(z)}{1+z}+f(z)\;\lambda''(z)+\frac{\Omega _{\Lambda } \,\lambda '(z)}{(1+z)^4\;
   \bar{ H} (z)\, \Omega _m}+\frac{\Omega _{\Lambda }\,
   \lambda (z)}{(1+z)^5 \,\Omega _m}=0\\\nonumber
  && r_1'(z)-\left(1+ \frac{3 \,H_0^2\, (1+z)^3 \,\Omega _m}{2\, \bar{  H}(z)^2}\right)\,
   \frac{\a_1(z)}{( 1+z)\, \bar{  H} (z)}-\\ &&\nonumber \qquad \qquad
   \frac{\Omega _{\Lambda } \,\left(\bar{H}(z)\, \lambda (z)+3\,
   (1+z) \,\lambda '(z)\right)}{3 \,(1+z)^5\, \bar{  H}(z)^2\,
   \Omega _m}+k(z)\,\lambda''(z)=0
      \eea
   where $\lambda(z)\equiv \lambda(r_0(z))$ and with initial conditions $\alpha_1( 0)=0, \;r_1( 0)=0$.
   Then, using (\ref{eqfrwR}),
   we  computed the various distances (\ref{da}) as a function of the redshift taking two different functions of the radial coordinate
   \bea\nonumber
   \Lambda(r)&=&3\,H_0^2\,\Omega_{\Lambda}\;\left(1+\lambda^{\mbox{\tiny(i)}}(r)\right ),\qquad i=1,2\\
   \label{lam1}
   \lambda^{\mbox{\tiny(1)}}(r)&=&\lambda_0\;e^{-\frac{(r-r_0)^2\,H_0^2}{\Delta^2}}\;\left(1-e^{-\frac{r^2\,H_0^2}{\Delta^2}}\right)
   \\ \label{lam2}
   \lambda^{\mbox{\tiny(2)}}(r)&=&\lambda_0\,\left(1-
   \,\left(\frac{1-\tanh[\frac{(r+r_0)\,H_0}{2 \,\Delta}]\,\tanh[\frac{(r-r_0)\,H_0}{2 \,\Delta}]}{1+\tanh[\frac{r_0\,H_0}{2 \,\Delta}]^2}\right)\right)
   \eea
 \begin{figure}[h! %htbp
  ]
        \centering
        \begin{minipage}[c]{.40\textwidth}
          \centering%\setlength{\captionmargin}{0pt}%
          \includegraphics[width=.80\textwidth]{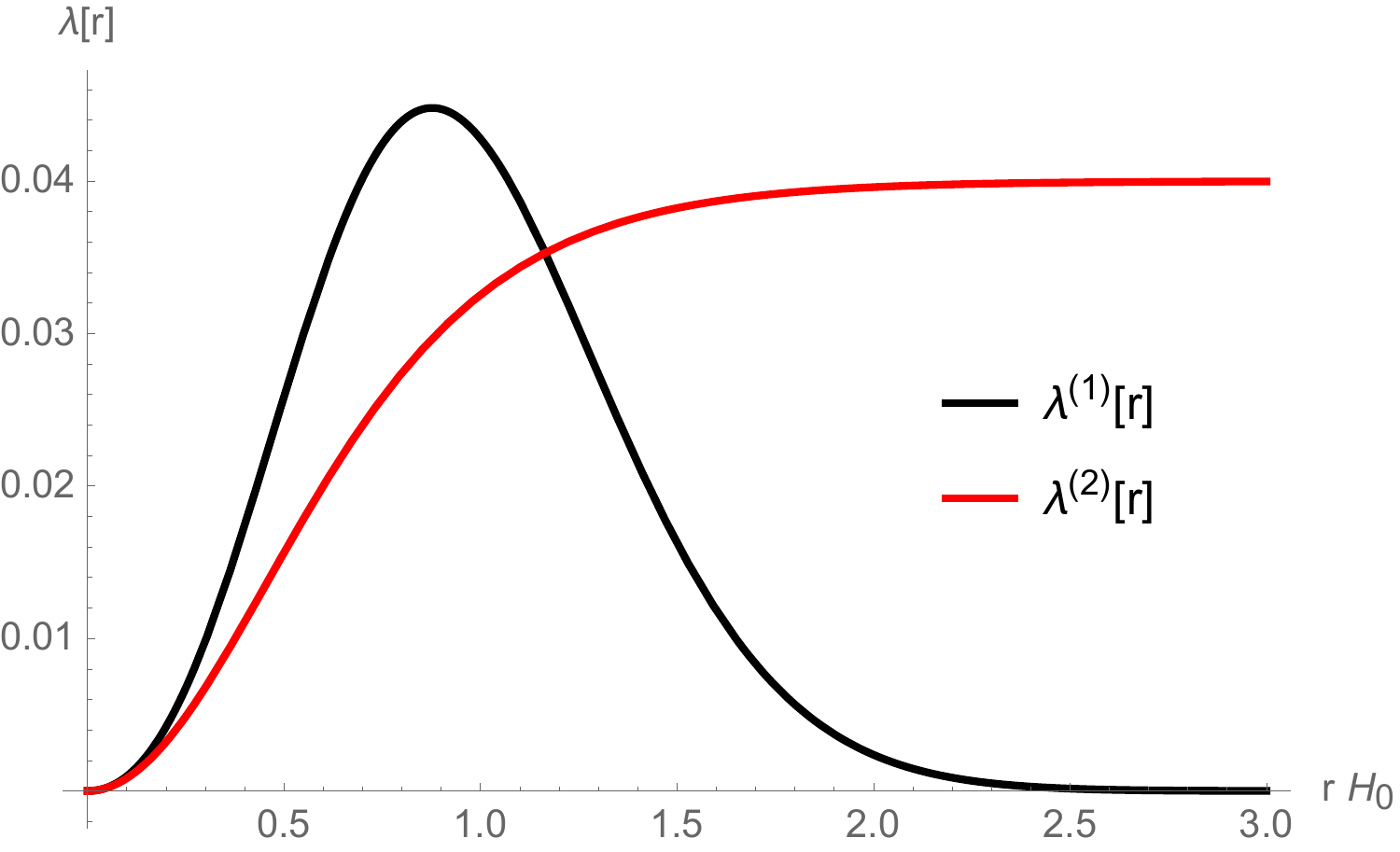}
          \caption{Plot of $\lambda^{(i)}(r)$ for the model (i=1) (\ref{lam1})  with coefficients given in (\ref{model1}) 
          and  the model (i=2) (\ref{lam2})   with coefficients given in (\ref{model2}).
          }
        \end{minipage}%
%        \caption{Didascalia comune alle
%          due figure\label{fig:minipage2}}
      \end{figure}
        In particular, for $\lambda^{\mbox{\tiny(1)}}(r)$ we have a bump at $r\sim r_0$, while for  $\lambda^{\mbox{\tiny(2)}}(r) $ the perturbation start from zero and becomes  a constant $ \lambda_0$ far away. 
 We imposed for both cases $\lambda'(r=0)=0$ (see \ref{l1r}). 
     The parameter 
  $\lambda_0$ characterises the size of the void, $r_0$ and $\Delta$ characterise  the transition to uniformity.
 \begin{figure}[%h! %htbp
  ]
        \centering
        \begin{minipage}[c]{.40\textwidth}
          \centering%\setlength{\captionmargin}{0pt}%
          \includegraphics[width=.80\textwidth]{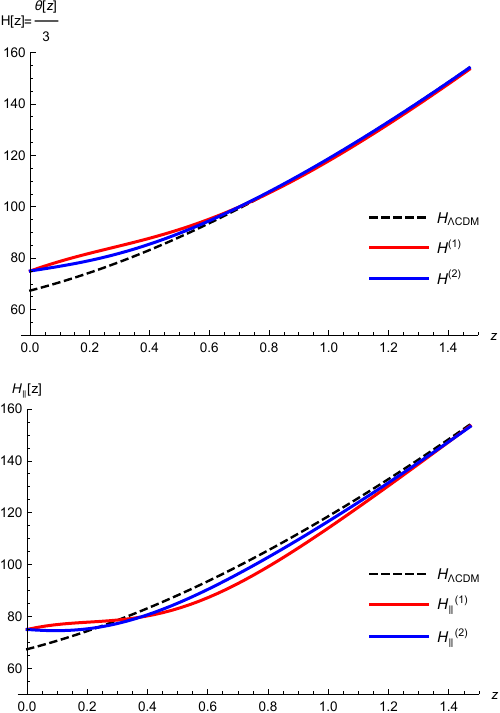}
          \caption{In the upper fig. we show the Hubble (volume expansion) parameter  ${\cal H}=\frac{\theta}{3}$ (\ref{Hvol})
          as a function of redshift for the   $\Lambda$CDM model ${\cal H}_{\Lambda CDM}$ and  for the models (1)  ${\cal H}^{\mbox{\tiny(1)}} $(\ref{lam1}) and (2) ${\cal H}^{\mbox{\tiny(2)}} $ (\ref{lam2}). 
          In the lower fig. we show the radial Hubble expansion rate (\ref{Hpar}, \ref{Hr}) always for the models (1)  ${\cal H}_{\parallel}^{\mbox{\tiny(1)}} $(\ref{lam1}) and (2) ${\cal H}_{\parallel}^{\mbox{\tiny(2)}} $ (\ref{lam2}). The Hubble scale is expressed  in ${\rm km \;s^{-1} \;Mpc^{-1}}$.}
        \end{minipage}%
        \hspace{10mm}%
        \begin{minipage}[c]{.40\textwidth}
          \centering%\setlength{\captionmargin}{0pt}%
          \includegraphics[width=.80\textwidth]{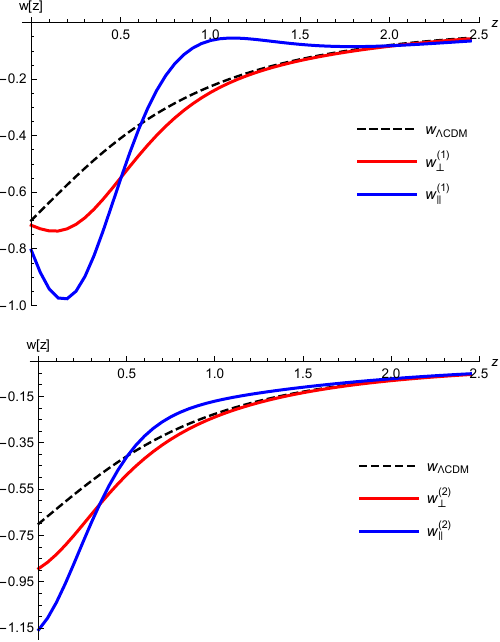}
          \caption{Effective equations of states relative to the FRW $\Lambda$CDM model $w_{\Lambda CDM}$, to the radial and transverse expansion rates $w_{\parallel/\perp}$ (\ref{wra}), for the models (1) (\ref{lam1}) and (2) (\ref{lam2}).}
        \end{minipage}
%        \caption{Didascalia comune alle
%          due figure\label{fig:minipage2}}
      \end{figure}
For the choice of the parameter space
 we impose the following constraints:
 \begin{itemize}
\item  Matching with the $\Lambda$CDM model in the early universe (as you can see from fig. 2, already present for $z\sim 1.5$), so we start with 
\bea\label{Hpl}
\left.{\cal H}_{\parallel}^{\mbox{\tiny(1,2)}}(z)\right|_{z\gg1}=
\left.{\cal H}_{\Lambda CDM}(z)\right|_{z\gg1} \Rightarrow%\quad
%\\
%\frac{H_0}{(1+z)^{3/2}}=\frac{ H_0^{CMB}}{(1+z)^{3/2}}\quad {\rm with}
 \quad H_0=H_0^{CMB}=67.4\;{\rm km\;s^{-1}\;Mpc^{-1}}
\eea
as given by (\ref{Hr}, \ref{Hcdm}, \ref{infr}). 
\item then we tune the parameters, for model (1) (\ref{lam1}) and model (2) (\ref{lam2})  
in such a way that $\lambda^{\mbox{\tiny(1)}''}\!(0)=\lambda^{\mbox{\tiny(2)}''}\!(0)$
%$\lambda^{\mbox{\tiny(1)}}''(0)=\lambda^{\mbox{\tiny(2)}}''(0)$
in order to get the same value at $r=0$
\bea
{\cal H}^{\mbox{\tiny(1)}}_{\parallel}(z=0)={\cal H}^{\mbox{\tiny(2)}}_{\parallel}(z=0)=H_0\;(1+0.667\;\lambda''(0))=75\;{\rm km\;s^{-1}\;Mpc^{-1}}
\eea
that implies
\bea \label{la2}
\lambda(0)=\lambda'(0)=0,\quad \lambda''(0)=0.17
\eea
\end{itemize}
Note the relationship
\bea
\rho(t_0,r=0)=3\;H_0^2\;\left(1-0.074\;\lambda''(0) \right)\rightarrow \frac{{\cal H}_{\parallel}(0)- H_0 }{H_0}\equiv \frac{\Delta H_0}{H_0}=-9.05\;\delta_0
\eea where $\delta_0=\Delta\rho(t_0,r=0)/\rho(t_0,r=0)$
to be compared with the expression valid in a FRW universe for  an adiabatic perturbation in density  of our local spacetime \cite{turner,marra}
\bea
 \frac{\Delta H_0}{H_0}=-\frac{1}{3}\;f(\Omega_m)\;\delta_0
\eea
where $f(\Omega_m)\sim0.5$. 
A $9\%$ of shift for the Hubble constant (as observed) in a $\Lambda$CDM model implies $\delta^{\Lambda CDM}_0\sim 0.5$ while in our 
N$\Lambda$CDM model  $\delta^{N\Lambda CDM}_0\sim 0.01$.
\\
For the nearby effective equation of state (\ref{wra}) 
and the effective acceleration (\ref{acc}), we also get
  \bea
  w_{\parallel}(0)=-0.7-2.695\;\lambda''(0)+0.445\;\lambda'''(0)
  \\
  w_{\perp}(0)=-0.7-1.140\;\lambda''(0)+0.222\;\lambda'''(0)
  \\
  q(0)=-0.55-4.043\;\lambda''(0)+0.667\;\lambda'''(0)
  \eea
 A particular realisation of eq. (\ref{la2}) in the  parameter space of 
 (\ref{lam1}) and (\ref{lam2}) is given by
  %\subsection{Geodesics for partially exactly solvable model }
  \bea\label{conf}
&& H_0=67.4\;{\rm km\;s^{-1}\;Mpc^{-1}},\; \Omega_m=0.3,\;\Omega_{\Lambda}=0.7,\;
  \\
  model \;(1)&:&\quad r_0=0.5/H_0,\;\Delta=0.8,\;\lambda_0=8\; 10^{-2} \label{model1}
  \\
  model \;(2)&:&\quad r_0=0.2/H_0,\;\Delta=0.328,\;\lambda_0=4\; 10^{-2} \label{model2}
  \eea
%  \bea
%  {\cal H}_{\parallel}(0)^{(1)}={\cal H}_{\parallel}(0)^{(2)}=75\;{\rm km\;s^{-1}\;Mpc^{-1}}
%  \eea
  as a result $\lambda^{\mbox{\tiny(1)}''' }(0)=0.793,\;\;\lambda^{\mbox{\tiny(2)}''' }(0)=0$ 
  so   that   we get
  \bea
  w_{\parallel}^{ \mbox{\tiny(1)} }(0)=-0.80,\quad
  w_{\perp}^{ \mbox{\tiny(1)}}(0)=-0.72,
  \quad q^{ \mbox{\tiny(1)}}(0)=-0.88,\\
  w_{\parallel}^{ \mbox{\tiny(2)}}(0)=-1.16,\quad
  w_{\perp}^{\mbox{\tiny(2)}}(0)=-0.89,
  \quad q^{\mbox{\tiny(2)}}(0)= -1.09
  \eea
  Note however that the full $z$ dependence of the above observables results quit not trivial as shown in figure 3 .
Comparing with the prediction of   $q_0=-1.08\pm 0.29$ \cite{Camarena:2019moy},
% where the  estimations only use supernovae in the redshift range $0.023\leq z\leq0.15$
 we see an interesting agreement with data (the local determination of the deceleration parameter for the standard $\Lambda$CDM model  gives $q=-0.55$ for $\Omega_m=0.3$).
  For the future it will be interesting to provide a functional analysis that produces a  best fit model for
     these observables.
   Here   we have given a  simple realisation of the  model's potential.

  \subsection{Geodesics for small $r$ expansion}
  \label{Gr}
  In the small redshift expansion from eqs (\ref{eqzt}, \ref{dzt}) we get
   \bea
   a(z)&=&1-z+\left(1-\frac{3}{4}\,\Omega_{\chi}+\frac{R_2'(t_0)-R_2(t_0)}{2\,{\cal H}_0} \right)\\
   t(z)&=&t_0-\frac{z}{{\cal H}_0}+\left(  \frac{1}{2}-\Omega_{\chi}-\frac{3\;\Omega
   _{\mathit{m}}}{4}+\frac{R_2'(t_0)}{2\;{\cal H}_0^2}-\frac{R_2(t_0)}{2\;{\cal H}_0}
   \right)\frac{z^2}{{\cal H}_0}+{\cal O}(z^2),\\
    r(z)&=&\frac{z}{{\cal H}_0}+\left( \Omega_{\chi}-\frac{3\;
   \Omega _{\mathit{m}}}{4}-\frac{R_2'(t_0)}{2\;{\cal H}_0^2}
   \right)\frac{z^2}{{\cal H}_0}+{\cal O}(z^2)
   \eea
    (where we used  $\Omega_\Lambda=1-\Omega_m-\Omega_{\chi}$).
  We can parametrize the various corrections as functions of the expansion coefficients of the acceleration and shear functions, see eq. (\ref{RAS}):
   \bea
   {\cal A}(z)&=&{\cal H}_0\left({\cal A}_1\,z+{\cal A}_2\,\frac{z^2}{2}+...\right),
   \qquad
     {\Sigma}(z)={\cal H}_0\left({\Sigma}_1\,z+{\Sigma}_2\,\frac{z^2}{2}+...\right)
   \eea
   where ${\cal A}_i$ and ${\Sigma}_i$ are constants depending from the initial conditions and free parameters:
   \bea\label{locA}
  {\cal A}_1&=& \frac{3}{2} \;\Omega_{\chi},
  \qquad  
  {\Sigma}_1  = \frac{1}{3\,{ \cal H}_0}\;\left(R_2'(t_0)-R_2(t_0) \right)
  \eea
  \bea
  {\cal A}_2&=&\frac{3 \, \mathcal{A}_1  \, \Sigma
   _1}{2}+\frac{\mathcal{A}_1^2}{3}+\frac{F_3\left(t_0\right)}{4 \, 
   \mathcal{H}_0^3}+\mathcal{A}_1 \left(-\frac{3 \,  \Omega _m}{8}-1\right)-\frac{3 \, 
   \mathcal{A}_1  \, R_2'\left(t_0\right)}{4 \,\mathcal{H}_0}
    \\\nonumber
     {\Sigma}_2
    &=& -\frac{\mathcal{A}_1}{3}+\Sigma _1 \left(\frac{\mathcal{A}_1}{6}+\frac{1}{8} \left(3 \, 
   \Omega _m+4\right)\right)+\frac{R_3'\left(t_0\right)-Y_3(t_0)}{9 \, 
   \mathcal{H}_0^2}-\frac{\Sigma _1  \, R_2'\left(t_0\right)+\frac{1}{6} \, 
   R_2''\left(t_0\right)}{\mathcal{H}_0}+\frac{9  \, \Sigma _1^2}{4}
       \eea
     Note that $R_2'\left(t_0\right)$ (and also $R_2''\left(t_0\right),\;R_3'\left(t_0\right)$) can be reads from eq. (\ref{R2}) as a function of the initial conditions $R_2\left(t_0\right)$ and the various parameters.    From (\ref{da}), the luminosity distance - redshift relationship can be parametrised in the usual way as  
   \bea\label{dec}
   d_L(z)&\equiv&
   \frac{z}{{\cal H}_0}+\left(1-Q_0
   \right)\frac{z^2}{2\;{\cal H}_0}+{\cal O}(z^3)
%   \\
%   &=&\frac{z}{{\cal H}_0}+\left(1+ \Omega_{\chi}-\frac{3\;
%   \Omega _{\mathit{m}}}{4}-\frac{R_2'(t_0)}{2\;{\cal H}_0^2}+\frac{R_2(t_0)}{2\;{\cal H}_0}
%   \right)\frac{z^2}{{\cal H}_0}+{\cal O}(z^2)
   \eea
    where we observe as  the deceleration parameter $Q_0$, deduced by $d_L$,
     is different from the usual dynamical deceleration parameter $q_0=-\left. \frac{a''\;a}{(a')^2}\right |_{t_0}$ \cite{sanchez}: 
    %(we parametrise the acceleration and the shear, at leading order, as
      \bea
   Q_0&=& -1+\frac{3}{2}\;\Omega_m-\frac{4}{3}\;{\cal A}_1+ 3\; \Sigma_1  ,
  \qquad   q_0=
   -1+\frac{3}{2}\;\Omega_m-\frac{ {\cal A}_1}{ 3 },\;\;
   \\&&
  \qquad\qquad\qquad Q_0-q_0=%-\frac{3\;\Omega_X}{2}+\frac{Y_2'(t_0)}{ {\cal H}_0^2}-\frac{Y_2(t_0)}{ {\cal H}_0}=
   - {\cal A}_1 + 3\;{\Sigma_0}  \nonumber
   \eea
   The radial Hubble parameter results
   \bea
   {\cal H}_{\parallel}(z)&=&
   \mathcal{H}_0+  \mathcal{H}_0 \left(-\frac{4  \, \mathcal{A}_1}{3}+\frac{3  \, \Omega _m}{2}+3 \,  \Sigma
   _1\right) \, z+\\&&\nonumber
  z^2 \left(-\frac{1}{4}  \, R_2''\left(t_0\right) +{\cal H}_0\, \left(-\frac{2\, \mathcal{A}_1^2}{9}-2\, \mathcal{A}_2+\Sigma _1 \,\left(\frac{3\,
   \mathcal{A}_1}{4}+\frac{3}{16}\, \left(4-9 \,\Omega _m\right)\right)+
   \right.\right. \\&& \left.\left. \mathcal{A}_1\,
   \left(\frac{5 \,\Omega _m}{4}+\frac{1}{3}\right)+\frac{3}{8} \left(4-3 \,\Omega _m\right)\,
   \Omega _m-\frac{9 \,\Sigma _1^2}{8}+\frac{9 \,\Sigma _2}{2}\right)
  \right) +... \nonumber \eea
   and the difference with the orthogonal Hubble rate (\ref{Hper}) is
   \be
  {\cal H}_{\parallel}(z)-{\cal H}_{\perp}(z)= {\cal H}_0 \left(z \,\left(\frac{3 \,\Sigma
   _1}{2}-\mathcal{A}_1\right)+z^2 \,\left(3 \,\Sigma _2-2 \,\mathcal{A}_2\right)+...\right)
   \ee
   Finally the  effective radial and angular equations of state (\ref{wra})  at leading order ${\cal O}(z)$ result 
   \bea
    w_{\perp}(z)&=&-1+\Omega_m-\frac{2}{9}\;{\cal A}_1+\Sigma_1+...,\\
    w_{\parallel}(z)&=&-1+\Omega_m-\frac{8}{9}\;{\cal A}_1+2\;\Sigma_1+ \nonumber
 \left(-\frac{40 \,
   \mathcal{A}_1^2}{27}-\frac{8 \,
   \mathcal{A}_2}{3}+\mathcal{A}_1
   \left(\frac{1}{9} \left(39  \,\Omega
   _m-4\right)+\frac{19 \, \Sigma
   _1}{3}\right)+
    \right.\\&&\left.
    \Sigma _1
   \left(3-\frac{33  \,\Omega
   _m}{4}\right)-3  \,\Omega _m^2+3 \,
   \Omega
   _m-\frac{R_2''\left(t_0\right)}{3\,
   \mathcal{H}_0}-\frac{15 \, \Sigma
   _1^2}{2}+6 \, \Sigma _2\right)\,z+...
   \eea
   Comparing with the prediction of  \cite{Camarena:2019moy}, where the  estimations only use supernovae in the redshift range $0.023\leq z\leq0.15$, that gives
   ${ \cal H}_0=75.35\pm 1.68\;km\,s^{-1}\,Mpc^{-1}$ and $Q_0=-1.08\pm0.29$ \footnote{Note that the local determination of the deceleration parameter for the standard $\Lambda$CDM model  gives $q=-0.55$ for $\Omega_m=0.3$.}
%we get
%$q=-0.91$ for   model  1 and $q=-1.09$ for   model  2
%%.to compare with $q=-1.08\pm 0.29$.
%
% From  \cite{Camarena:2019moy} 
 we can estimate the nearby equation of state, the acceleration and the $\Omega_\chi$ parameter to
   \bea \nonumber
% \!\!\!  && { \cal H}_0=75.35\pm 1.68\;km\,s^{-1}\,Mpc^{-1}
%   \\\nonumber
%  && Q_0=-1.08\pm0.29\\\nonumber
   && w_{\parallel}=-1.05\pm 0.19+\left( 6 \,
   \Sigma _2-\frac{8\,
   \mathcal{A}_2}{3}-\frac{Y_2''(t_0)}{
   3  \,{\cal H}_0}-\frac{3\, \Sigma _1^2}{4}+(0.73\pm0.14)+(2.32\pm0.07)\,\Sigma_1\right)z
  \\\label{Hnear}
 && {\cal A}_1=0.40\pm0.22+\frac{9}{4}\,\Sigma_1,\qquad
  \Omega_{\chi}=0.27\pm 0.15+\frac{3}{2}\,\Sigma_1
   \eea
   To disentangle the degeneracy ${\cal A}_1-\Sigma_1$ we need an extra independent cosmographic measure that can be the   redshift drift, as suggested in \cite{Partovi:1982cg}.
 
  \section{Conclusions}
 We analysed an adiabatic perfect fluid cosmological model characterised by
 two terms into the Lagrangian  (\ref{lagU}), one featuring a CC and another a DM component, both terms are builded with the same field  content (goldstone modes of the spontaneously broken space and time) (\ref{fields}). 
 In this sense we have a single fluid mimicking both  DE and DM. 
% His thermodynamical properties are quite unusual, corresponding to an unstable fluid.
  %His adiabatic behaviours are manifest in a
  The presence of a   space dependent CC results thermodynamically related to the conserved (in time)  entropy density (\ref{as}).
  Such a   DE component breaks  the homogeneity of the space time and induces
  a non trivial space dependence for the pressure (\ref{par}) such that the
   comoving observers are not anymore geodesic ${\cal A}_\mu\neq0$ (\ref{as}).
\\
  We study the equation of motion of such a system using a 1+1+2 formalism where we use two 
    background vectors: one is time-like $u^\mu$  (the geodesic of the fluid) and the other is space-like   $v^\mu$ (proportional to the non zero  acceleration) (\ref{accv}).
  \\
  In chapter  (\ref{ch:112})  we give all the covariant equation for a LRS space  (\ref{eqA},
  \ref{pL}, \ref{eqe}, \ref{eqp}, \ref{eqpe}) and part of some non perturbative solutions
  (\ref{kf}).
  \\
   In chapter (\ref{ng}) we give the eqs for the propagation of light bundles  and we stress   the difference in between the  radial ${\cal H}_{\parallel}$ and transverse ${\cal H}_{\perp}$ expansion rates (\ref{Hpar}, \ref{Hper}).
  \\
  In chapter (\ref{ca}) we study the eqs of motion  in a comoving coordinate system using a rotationally invariant  Lema$\hat{\rm i}$tre metric (\ref{g}) characterized by three form factors functions of the $t,\,r$ variables $(F(t,r),\,X(t,r),\,R(t,r))$. The  eqs of motion (\ref{eqNM}) are solved in two limiting cases.
  \\
   In chapter (\ref{chFRW}) we perturb a FRW background with a small $r$-dependent cosmological constant (\ref{eqll}) while in chapter (\ref{Gfrw}) we give the corresponding solutions of the null geodesics.
   \\
  In chapter (\ref{chY}) instead we study the solutions to (\ref{eqNM}) in the  small $r\,H_0\ll1$ expansion regime and in chapter (\ref{Gr}) we give the  corresponding null geodesic paths. 
  \\
  As observable, to check the outcome of our model,  we study the luminosity distance $d_L$ (\ref{da}), the effective eqs of state 
  along the radial and transverse direction $w_{\parallel,\perp}$ (\ref{wra}) computed along the light of sight.  In chapter (\ref{Gfrw}) we analyse two kinds of CC $r$-profiles (\ref{lam1},\ref{lam2}) and we report our main results in fig. 2 and 3.
  \\
  In chapter
  (\ref{Gr}) the small $r$  expansion is translated in a small $z$ redshift expansion where we can extract 
    the deceleration parameter $Q_0$ (\ref{dec}) as a function of constants describing the background evolution.    It is interesting to note as, in such a limit, the background Hubble  law (\ref{Hrt}) shows the presence of an extra matter component $\Omega_\chi$ (proportional to the local acceleration parameter (\ref{locA})) whose equation of state result $w_\chi=-4/3$.
    \\
   %At the light of the present discussions on the status of
   To simplify our computations we put the observer at the center of the void region
   violating the Copernican (Cosmological) Principle.
%    that states that we do not occupy a special location in the Universe.
    \footnote{Many anomalies afflict the present FRW representation of the universe,  see \cite{cea} for a review.   }  
   Our model represent an alternative to the more popular  spherically symmetric LTB model where the 
   matter component of the universe is the responsible of the inhomogeneity.
 Assuming  the $\Lambda$CDM model,
 at present, the discrepancy between the Hubble values measured around local distance ladders and from CMB Planck data  (see \cite{DiValentino:2021izs}, \cite{Hu:2023jqc} for a full reference list) has reached the 4-6 $\sigma$ level.
This model represents a minimal modification to the $\Lambda$CDM scenario (for this reason we 
%that we
  named next to minimal  $\Lambda$CDM model (N$\Lambda$CDM)) able to  modify the null geodesic paths in agreement with present data.
  This paper is dedicated to the theoretical analysis  of the model therefore we have not carried out a statistical analysis for the best possible parameter space of the model by just analysed two generic possible space dependent shapes for the CC shown in fig.1 
  The first kind of shape is featuring a bump of $\Lambda(r)$, 
  while the second one a smooth  increase of $\Lambda(r)$ till a constant value.
  Choosing appropriate parameters we first adjust the discrepancy for the Hubble parameter nearby and far away (Planck), then we cheque  the implications for the effective equations of state.
    For example we note that data favours  the second model where an effective nearby equation of state $w\leq-1$ is obtained mimicking a phantom dark energy component  \cite{carroll} \cite{Lee:2022cyh}. 
%  Matching the Hubble parameter to the CMB value far away from the center we  find the second shape (\ref{lam2})   is preferred by the observations.
Following the classification mechanism solutions for the $H_0$-crisis \cite{DiValentino:2021izs},
we see that this kind of  models can be inserted in the class of the alternative proposals operating at late time: {\it Local Inhomogeneity}.
 It is clear that to probe this kind of  radial inhomogeneity around us we further have to do many other  test:
  CMB spectral distortions,  BAO, type Ia supernovae, cosmic chronometers, ect. 
as is in progress  for the $\Lambda$CDM model  \cite{Perivolaropoulos:2021jda}.

\section*{Acknowledgments}
I have to deeply thank M. Celoria, L. Pilo and R. Rollo for many 
 %($almost\, endless$ $\ddot\smile$) 
 discussions on the subject.
%$''${\it E il giorno della fine non ci servir\u a l' inglese}$''$ from F. Battiato.
 %in the album {\it Il Re Del Mondo}

\appendix
\section{ {1+3} and 1+1+2 Formalism}

\begin{itemize}
\item{1+3} Formalism
\\
Projection tensor on the metric of the 3-space (S) orthogonal to $u^\a$
\bea
h_{\a\b}=g_{\a\b}+u_{\a}\;u_{\b},\qquad h^\a_\a=3
\eea
Any 3-vector, $V^\a$, can be irreducibly split into a component along $u^\a$ and a sheet
component ${\cal V}^\a$, orthogonal to  $u^\a$
\be
V^\a=V\;u^\a+{\cal V}^\a,\qquad V=-V_\a\;u^\a,\quad {\cal V}^\a=h^\a_\b\;V^\b
\ee
The covariant derivative of a scalar function $f$ projected along $u^\a$ is so defined
\bea
%\hat f=n^\a\nabla_\a f,\qquad 
\dot f=u^\a\nabla_\a f
\eea
We are now able to decompose the covariant derivative of $u^\a$ orthogonal to $u^\a$ giving
\be
\nabla_\a\;u_\b=-u_\a \;{\cal A}_\b+\frac{1}{3}\;\theta\; h_{\a\b}+\sigma_{\a\b}+\omega_{\b\a}
\ee
\begin{itemize}
\item 4-Acceleration
\be
{\cal A}_\a\equiv u^\b\nabla_\b\;u_\a\equiv \dot u_\a 
\ee
\item Expansion
\be
  \theta \equiv \nabla^\a\;u_\a
 \ee
 \item Shear
 \be
 \sigma_{\a\b}=h^\lambda_\b\;\nabla_{(\lambda}\;u_{\rho )}\;h_\a^\rho-\frac{1}{3}\;\theta\; h_{\a\b} 
 \ee
 \item Vorticity
 \be
 \omega_{\a\b}=h^\lambda_\b\;\nabla_{[\lambda}\;u_{\rho ]}\;h_\a^\rho,\qquad \omega^\a=\e^{\a\b\g}\;\omega_{\b\g}
\ee
\end{itemize}
The decomposition defined above is also used to define a set of derivatives operators acting on various kind of tensors:
 \bea
 &&\dot X^{\a..\b}_{\gamma..\delta}=u^\lambda\;\nabla_\lambda\;X^{\a..\b}_{\gamma..\delta}\\
 && D_\lambda X^{\a..\b}_{\gamma..\delta}= h^{\lambda_1}_{\lambda}\;
 h^{\a}_{\a_1}\;h^{\b}_{\b_1}\;...\;h^{\gamma}_{\gamma_1}\;h^{\delta}_{\delta_1}\;\nabla_{\lambda_1}\;X^{\a_1..\b_1}_{\gamma_1 ..\delta_1}
 \eea
 
 \item {1+1+2} Formalism
 \\
 The 1 + 1 + 2 approach is based on a double foliation of the spacetime: a given manifold is first foliated in spacelike 3-surfaces and then these 3-surfaces are foliated in 2 surfaces. These foliations are obtained by defining a congruence of integral curves of the time-like ($u^\a\;u_\a = - 1$) vector field  $u^\a$  and a congruence of integral curves defined by the spacelike ($v^\a\;v_\a =   1$) vector $v^\a$.
 \\
 Projection tensor which represents the metric of the 2-spaces (W) orthogonal to $u^\a$ and $v^\a$.
\bea
N_{\a\b}=h_{\a\b}-v_{\a}\;v_{\b},\qquad N^\a_\a=2
\eea
Any 3-vector  $V^\a$  can be irreducibly split into a component along $v^\a$ and a sheet
component ${\bf V}^\a$, orthogonal to  $v^\a$ and $u^\a$.
\be
V^\a=V\;v^\a+{\bf V}^\a,\qquad V=V_\a\;v^\a,\quad {\bf V}^\a=N^\a_\b\;V^\b
\ee
The decomposition defined above can be also used to define a set of derivatives operators:
 \bea
&&  \hat X^{\a..\b}_{\gamma..\delta}=v^\lambda\;\nabla_\lambda\;X^{\a..\b}_{\gamma..\delta}\\
 && \delta_\lambda X^{\a..\b}_{\gamma..\delta}= N^{\lambda_1}_{\lambda}\;
 N^{\a}_{\a_1}\;h^{\b}_{\b_1}\;...\;N^{\gamma}_{\gamma_1}\;N^{\delta}_{\delta_1}\;\nabla_{\lambda_1}\;X^{\a_1..\b_1}_{\gamma_1 ..\delta_1}
 \eea
  The hat-derivative is the derivative along the vector field $v^\a$ in the surfaces orthogonal to $u^\a$, for a scalar function $f$ we have the definitions
\bea
\hat f=v^\a\nabla_\a f %,\qquad \dot f=u^\a\nabla_\a f
\eea
For the mixed double derivative we have the relationship
\bea\label{mix}
\hat{\dot{f}}-\dot{\hat{f}}=-{\cal A} \;\dot f+\left(\frac{\theta}{3}+\Sigma\right)\;\hat f
\eea
We are now able to decompose the covariant derivative of $v^\a$ orthogonal to $u^\a$ giving
\bea
\nabla_\a\;u_\b=-{\cal A}\;u_\a\;e_\b+\left(\frac{\theta}{3}+\Sigma\right)\;v_\a\;v_\b+
\left(\frac{\theta}{3}-\frac{\Sigma}{2}\right)\;N_{\a\b}+\Omega\;\e_{\a\b}
\\
\nabla_\a\;v_\b=-{\cal A}\;u_\a\;u_\b+\left(\frac{\theta}{3}+\Sigma\right)\;v_\a\;u_\b+
 \frac{\phi}{2} \;N_{\a\b}+\xi\;\e_{\a\b}
\eea
\be
D_\a\;v_\b=v_\b \;a_\b+\frac{1}{2}\;\phi\; N_{\a\b}+\xi\;\e_{\a\b}+\zeta_{\a\b}
\ee
where
\begin{itemize}
\item Sheet expansion
\be
\phi=\delta_\a\;v^\a=D_\a\;v^\a
\ee
\item Twisting of the sheet (the rotation of $v^\a$)
\be
\xi=\frac{1}{2}\e^{\a\b}\;\delta_\a v_\b
\ee
\item Acceleration  of $v^\a$
\be a_\a=\hat v_\a
\ee
\item Shear of $v^\a$
\be
\zeta_{\a\b}=\delta_{\{\a}\;v_{\b\}}
\ee

\end{itemize}
%EoM
%  \bea
%  \dot Z&=&0\to
%  \dot \sigma=0
%    \\ \dot b&+&\theta\;b=0,
%  \\ D_\mu p&=&-Z\; D_\mu \sigma  \\
%  (\rho+p)\;a_\mu&+&D_\mu p=0\to \;a_\mu=\frac{Z}{\Omega_m\;b}\; D_\mu \sigma       
%  \eea
%  \be
%  A=\frac{Z}{\Omega_m\;b}\; D_\mu \sigma \;v^\m=\frac{Z}{\Omega_m\;b}\;  \sqrt{(D\sigma)^2}
%  \ee
Note the following identities/definitions  
  \bea
  %&&\nabla_\a\;u_\b=\sigma_{\a\b}+\o_{\b\a}+\frac{\theta}{3}\;h_{\a\b}-u_\a\;{\cal A}_\b\\
&&   v\cdot u=0,\quad v^2=1,\quad  v\cdot a=-u\cdot \dot v,\quad v\cdot\dot v=0
\\
  && {\cal A}={\cal A}^\a  v_\a =-u^\a\;\dot v_\a ,\quad \Sigma=v^\a\sigma_{\a\b} v^\b,\quad  
  \Omega=v^\a \o_{\a\b} v^\b,\quad\\
  &&
  {\cal E}=v^\a E_{\a\b} v^\b,\quad
  {\cal H}=v^\a H_{\a\b} v^\b,\quad
     \eea
 where $E_{\a\b}$ and $H_{\a\b}$ are the electric and the magnetic part of
the Weyl tensor respectively.

\end{itemize}

\section{Thermodynamical stability}
\label{ch:stab}
 Stable equilibrium is constraining the second  order variations of the thermodynamical potentials, specifically we have\footnote{ We defined: $N=n\;V,\;S=s\;V,\;E=\rho \;V$ where $V$ is the volume}
 \be
 \delta^2 S\leq 0\quad {\rm or} \quad \delta^2 E \geq 0%,\qquad S=V\,s, \;\;E=V\,\rho
 \ee
 For  a more formal contact with the thermodynamical formalism we can define in a field theoretical framework some classical text book quantities \cite{callen}:   
 \begin{itemize}
 \item Specific heat at constant volume
 \be\label{cv}
 C_V=T\;\left.\frac{\partial S}{\partial T}\right |_{V,N}=V\;T\;U_{Y^2}
 \ee
 
 \item
 Specific heat at constant pressure
 \be
 C_p=T\;\left.\frac{\partial S}{\partial T}\right |_{p,N}=V\;T\;\left(U_{Y^2}-
 \frac{(b\;U_{bY}-U_Y)^2}{b^2\;U_{b^2}}\right)
 \ee
 \item
 Coefficient of thermal expansion
 \be
 \alpha=\frac{1}{V}\;\left.\frac{\partial V}{\partial T}\right |_{p,N}= 
 sign[T]\;\frac{(b\;U_{bY}-U_Y)}{b^2\;U_{b^2} } 
 \ee
 \item
 Coefficient of isothermal compressibility
 \be\label{kt}
 \k_T=-\frac{1}{V}\;\left.\frac{\partial V}{\partial p}\right |_{T,N}= 
 -\frac{1}{b^2\;U_{b^2}} 
 \ee
 \item
 Coefficient of adiabatic compressibility
 \be
 \k_S=-\frac{1}{V}\;\left.\frac{\partial V}{\partial p}\right |_{S,N}= 
 -\frac{U_{Y^2} }{b^2\;U_{b^2}\;U_{Y^2}- (b\;U_{bY}-U_Y)^2} 
 \ee
 
 \end{itemize}
 All of them satisfy the above   relations
 \be
 c_p-c_V=\frac{T\;\a^2}{\k_T},\qquad c_p\;\k_S=c_V\;\k_T
 \ee
 where $C_{p,V}\equiv V\;c_{p,V}$.
Working with densities,  thermodynamical stability requires the  $|s_{ij}|$ matrix 
\bea\label{smat}
s_{ij}=\begin{pmatrix} 
  \frac{\partial^2 s}{\partial \rho^2}&\frac{\partial^2 s}{\partial \rho\partial n} \\
  \frac{\partial^2 s}{\partial \rho\partial n}&\frac{\partial^2 s}{\partial n^2} 
\end{pmatrix}\eea
to be negative defined so that
\bea 
&& s_{\rho\rho}=-\frac{1}{T^2\,c_V}<0,\quad \det |s_{ij}|=\frac{1}{n^2\;T^3\,c_V\;\k_T}>0
\\
&& {\rm or}\quad
 c_V>0\qquad \& \qquad T\;\k_T>0
 %\qquad  \rightarrow\qquad  T\;\k_T\;c_V>0
 \eea
 and also  $c_p>c_V>0$ or  $T\;\k_T\;c_V>0$ bringing to the following constraint for the potential
 \bea\label{U2}
  \frac{ U_{Y^2}}{ U_{b^2}}<0
 \eea
 Note that such a conditions are non perturbative and background independent.\footnote{In ref. \cite{Ballesteros:2016kdx} it was studied only the case for $T>0$.}

  \section{  Cosmological Constant and $\Lambda$-Media
   }\label{ch:CC}
  
The largest symmetry inside the internal scalar space is a full
unitary Diffeomorphism transformation
\be
\Phi^A\to f^A(\Phi^B)\;\;\;{\rm  where}\;\;\;\det \left |\frac{\partial f^A}{\partial\Phi^B}\right |=1
\ee 
whose feature is to select one specific invariant operator $Z$ that results the building block 
for a thermodynamical  non trivial CC: 
\bea
Z\equiv b\;Y=\frac{det |\Phi^A_\a |}{\sqrt{g}}=\frac{e^{\a\b\gamma\delta}}{\sqrt{g}}\;e_{A B C D}\;\Phi^A_{\a}
\;\Phi^B_{\b}\;\Phi^C_{\gamma}\;\Phi^D_{\delta}%\equiv \frac{\partial_{\a}\;J^\a}{\sqrt{g}}
\eea
where $e^{\a\b\gamma\delta}$ and $e_{A B C D}$ are the permutation tensors in the 4-Dim space-time and the 4-Dim scalar space.
\\
The action  generating an  effective CC perfect fluid results (that we named $\Lambda$-Media in \cite{Ballesteros:2016gwc,Celoria:2017idi})
\bea\label{flL}
S_{\Lambda}=\int d^4x \;\sqrt{g}\;\;U(Z)
\eea
The eqs of motion for the $\Phi^A$ scalar fields  (for $U''\neq 0$) are given by
\bea
&&\nabla_\mu \;det|\Phi^A_\a|=0 \quad \to\quad
\Psi_A^\a\;  \Phi^A_{\a\mu}=0 \;\;\;{\rm where}\;\;\;\Psi_A^\a=\frac{\partial x^\a }{\partial \Phi^A}
\eea
($\Psi_A^\a\;\Phi^A_\beta=\delta^\a_\beta,\;\;\Psi_A^\a\;\Phi^B_\a=\delta^A_B$) 
and imply the exact non perturbative solution
\bea
\nabla_\mu \;Z=0\quad \to\quad Z=Z_0\;\;{\rm constant \;in \;space\; and\; time}
\eea
The corresponding Energy Momentum Tensor (EMT)   can be easily obtained \footnote{We defined $U'=\partial U/\partial Z$ etc.}
\bea
T_{\mu\nu}=(U-Z\;U')\;g_{\mu\nu}\equiv -\Lambda \;g_{\m\n}
\eea
%Note that the EMT conservation implies, reproducing the result of the eqs of motion, the constancy of the product $b\;Y$:
%\bea
%\nabla^\nu\;T_{\mu\nu}=-Z\;U''\;\nabla_\mu Z=0
%\eea
so that we have an effective CC: $\Lambda =(-U+Z\;U')$.
%Due to the fact that a vacuum energy system has the same EMT we call $\Lambda$ also vacuum energy. 
For such a model we can write the energy density $\rho$, the pressure $p$ and the entropy per particle $\sigma$ as \cite{Celoria:2017idi}
\bea\label{eqL}
&&\rho_{\Lambda}=-p_{\Lambda}=\Lambda,\;\;\; %-U+Z\;U'=-U+Z\;\sigma_{\Lambda}\\
\sigma_{\Lambda}=\frac{s_{\Lambda}}{b}=U'
\eea
Using the thermodynamical analysis of the previous chapter,   for 
 the specific potentials $U=U(b\;Y) $, we get   that {\it  Perfect Fluid $\Lambda$-media 
 are thermodynamically unstable}. In fact we get
 \bea
 c_V= T\;b^2\;U'',\qquad
 %\qquad%\\&&
 %c_P=  0
 %\qquad%\\&&
 %\alpha=\frac{1}{T},\qquad
 %\\&&
 \k_T=-\frac{1}{b^2\;T^2\;U''}%=-\frac{\alpha}{c_V}
 ,\qquad
 %\qquad%\\&&
 %\k_S=\frac{1}{0}\;\to\;\infty
% \eea
 %\be
%c_P\;\k_S=
T\;c_V\;\k_T =-1
 \eea
  violating eq.(\ref{U2}). 
  Note the paradox  that {\it dynamically } the system is frozen to and exact de Sitter phase
  %(see (\ref{eqL}) ) 
  with no runaway solutions for the energy/entropy observables.

\section{Mixed derivatives for  thermodynamical quantities}\label{mixed}
We give the mixed derivatives (\ref{mix}) for the thermodynamical quantities $\rho,\;p,\;\sigma$ for a perfect fluid
 \bea
&&\hat{\dot{\rho}}-\dot{\hat{\rho}}={\cal A} \;\theta\;( \rho+p)+\left(\frac{\theta}{3}+\Sigma\right)\,\hat \rho
\\&&
\hat{\dot{p}}-\dot{\hat{p}}=-{\cal A} \;\dot p-\left(\frac{\theta}{3}+\Sigma\right)\,{\cal A}\;(\rho+p)\\
%\hat{\dot{\sigma}}-
&&\dot{\hat{\sigma}}=%-A \;\dot \sigma+
-\left(\frac{\theta}{3}+\Sigma\right)\,\hat \sigma
\eea

\section{Evolution equation for the radial form factor in almost FRW}

In order to get a single evolution equation for the radial form factor of the metric $R(t,r)$ we
can insert eq. (\ref{eqYh}) in (\ref{eqY}) so we get  
  $F$ as a function of $R$
\bea
 %\dot Y^2&=&-1+\hat{Y}^2+\frac{\Lambda}{3}\;Y^2+ \frac{\bf m}{3\;Y}\\
  \label{eqFF}
   F^2 &=&\frac{9 \;\mathbf{n}^2  \;R \;
   \left(R^{(1,0)}\right)^2}{R \;
   \left( \mathbf{m}'+R^3 \;
   \Lambda '\right)^2+3 \;
   \mathbf{n}^2 \left( 
   \mathbf{m}+\Lambda   \;R^3-3 \;
   R\right)}
   \eea
   Finally to have a self contained eq. for $R$ we insert (\ref{eqFF}) inside (\ref{eqF})
    \bea
  &&\label{eqy11}
  \frac{R^{(1,1)}}{R^{(1,0)}}\; 
   \left(R \; \left( 
   \mathbf{m}'+R^3\;  \Lambda
   '\right)+\frac{3 \; \mathbf{n}^2\; 
   \left(   \mathbf{m}+\Lambda \; 
   R^3-3 \; R\right)}{ 
   \mathbf{m}'+R^3\;  \Lambda
   '}\right)-
   \\ \nonumber
   &&  \;\left(R^3 \;\Lambda''+ 
   \mathbf{m}''\right)\;R- \frac{
   3}{2} \;\mathbf{n}^2 +\frac{R \;   
   \mathbf{n}' \; \left(  
   \mathbf{m}'+R^3 \; \Lambda
   '\right)}{\mathbf{n}}+\\
  &&\nonumber
  R^{(0,1)}  \;
   \left(-6\; R^3\; \Lambda
   '-\frac{3\; \mathbf{n}^2
   \left(  \mathbf{m}'\;
   \left(\Lambda  \;R^3-1/2\;
   \mathbf{m}\right)+R^3 \;\Lambda
   '\; \left(15/6\; \mathbf{m}+4\;
   \Lambda  \;R^3-9\;
   R\right)\right)}{R\; \left( 
   \mathbf{m}'+R^3\; \Lambda
   '\right)^2}\right) 
  =0
    \eea
    
\section{Evolution equation for the radial form factors in small $r$ regime }
Evolution for the form factor $R_{2,3}(t)$ where $R(t,r)=a(t)\,r+R_2(t)\,\frac{r^2}{2}+R_3(t)\,
\frac{r^3}{6}+...$. We prefer to shift the derivation, from the variable $t$ to $a(t)$ (\ref{Hrt}) such that
$f'(t)=%a'(t)\;f'(a)=
a\;\bar{\cal H}(a)\;f'(a)$.
\bea  \label{R2}
R_2'(a)=\frac{{\cal H}_0^2 \,R_2(a) \, \left(2 \, a^3 \, \Omega _{\Lambda }+3 \, a^4  \,\Omega _{\chi }-\Omega
   _m\right)}{2  \,a^4 \, \bar{\cal H}(a)^2}+\frac{\mathbf{m}_4 \left(3 \, \Omega _m-4 \, a^4 \, \Omega _{\chi }\right)}{216  \,a^3 \, \bar{\cal H}(a)^2 \, \Omega
   _m}+\frac{a  \,\ell _3}{3 \, \bar{\cal H}(a)^2  \,\Omega _m}
\eea
\bea\nonumber
&R_3'(a)&=-\frac{{\cal H}_0^2  \,R_3(a) \left(-2  \,a^3  \,\Omega _{\Lambda }-3 \, a^4 \, \Omega _{\chi }+\Omega
   _m\right)}{2  \,a^4 \, \bar{\cal H}(a)^2}+
   \frac{{\cal H}_0^2  \,\ell _3  \,R_2(a) \left(a^3  \,\left(3 \, a \, \Omega _{\chi }+4 \, \Omega _{\Lambda
   }\right)+7  \,\Omega _m\right)}{4  \,a^3 \, \bar{\cal H}(a)^4 \, \Omega _m}
   +
   \\&&\nonumber
   \frac{{\cal H}_0^4  \,R_2(a){}^2  \,\left(3  \,a^7  \,\Omega _{\chi }  \,\left(3  \,a \, \Omega _{\chi }+4 \, \Omega
   _{\Lambda }\right)+6  \,a^3 \, \Omega _m \left(11  \,a \, \Omega _{\chi }+6 \, \Omega _{\Lambda
   }\right)+9 \, \Omega _m^2\right)}{16 \, a^8  \,\bar{\cal H}(a)^4}
   -\frac{a^2  \,\ell _3^2}{12 \, \bar{\cal H}(a)^4  \,\Omega _m^2}
   +
   \\&&\label{R3}
     \frac{1}{\bar{\cal H}(a)^2} \left(\frac{3}{4}  \,H_0^4 \, \Omega _{\chi } \left(a^3 \, \left(7 \, a \, \Omega _{\chi }+6 \, \Omega _{\Lambda
   }\right)+9 \, \Omega _m\right)+\frac{a  \,\ell _4}{3 \, \Omega _m}\right)
\eea
where we used  $\Lambda_i\equiv 3\;{ \cal H}_0^2\;{\ell}_i$.

\end{document}